\documentclass[%
 reprint,
superscriptaddress,
 amsmath,amssymb,
 aps,
 prc,
]{revtex4-2}

\usepackage{amsmath}
\usepackage{textgreek}
\usepackage[utf8x]{inputenc}
\RequirePackage[T1]{fontenc}
\usepackage{graphicx}
\usepackage{dcolumn}
\usepackage{bm}
\usepackage{bm}
\usepackage{isotope}
\RequirePackage[sort&compress]{natbib}
\usepackage{subcaption}
\usepackage{mathtools} 
\usepackage{comment}
\RequirePackage[colorlinks,citecolor=blue,urlcolor=blue,linkcolor=blue]{hyperref}
\usepackage{url}

\begin{document}

\preprint{APS/123-QED}

\title{Bayesian uncertainty quantification on nuclear level density data and their impact on $(p,\gamma)$ reactions of astrophysical interest}

\author{A.~Chalil}
\altaffiliation{Present address: CEA, DES, IRESNE, Nuclear Measurement
Laboratory, F-13108 Saint-Paul-lez-Durance, France;
achment.chalil@cea.fr}
\affiliation{Univ. Lyon, Univ. Claude Bernard Lyon 1, CNRS/IN2P3, IP2I Lyon, F-69622, Villeurbanne, France}%
\author{C.~Ducoin}%
\affiliation{Univ. Lyon, Univ. Claude Bernard Lyon 1, CNRS/IN2P3, IP2I Lyon, F-69622, Villeurbanne, France}%
\author{O.~Stézowski}%
\affiliation{Univ. Lyon, Univ. Claude Bernard Lyon 1, CNRS/IN2P3, IP2I Lyon, F-69622, Villeurbanne, France}%
\author{N.~Millard-Pinard}%
\affiliation{Univ. Lyon, Univ. Claude Bernard Lyon 1, CNRS/IN2P3, IP2I Lyon, F-69622, Villeurbanne, France}%
\author{J.~Dudouet}%
\affiliation{Univ. Lyon, Univ. Claude Bernard Lyon 1, CNRS/IN2P3, IP2I Lyon, F-69622, Villeurbanne, France}%
\author{Y.~Demane}%
\affiliation{Univ. Lyon, Univ. Claude Bernard Lyon 1, CNRS/IN2P3, IP2I Lyon, F-69622, Villeurbanne, France}%
\author{M.~Chamseddine}%
\affiliation{Univ. Lyon, Univ. Claude Bernard Lyon 1, CNRS/IN2P3, IP2I Lyon, F-69622, Villeurbanne, France}%


\begin{abstract}
The $p$ process nucleosynthesis is responsible for the synthesis of 35 neutron-deficient nuclei from \isotope[35]{Se} to \isotope[196]{Hg}. An important input that can affect the modeling of this process is the nuclear level density at the relevant excitation energies of the nuclei involved in the reaction network. The OSLO method has been extensively used for the measurement of level densities in excitation energies of several MeV. In this work, Bayesian optimization has been used in order to estimate the 95\%  high density intervals for the parameters of two level density models optimized on the OSLO data. These uncertainties are then propagated on the cross sections of ($p,\gamma$) reactions leading to the compound nuclei \isotope[105,106]{Pd} and \isotope[105,106]{Cd} inside the astrophysically relevant energy range. Imposing constraints in this region of the isotopic chart is important for network calculations involving the nearby $p$ nuclei \isotope[102]{Pd} and \isotope[106]{Cd}. We discuss the reduction of the range of cross sections due to the uncertainties arising from the level density data compared to the range of the six default level density models available in TALYS and we highlight the need for level density data inside the astrophysically relevant energy ranges.
\end{abstract}
\maketitle


\section{Introduction}
\label{intro}

35 neutron-deficient nuclei, from \isotope[94]{}{Se} to \isotope[196]{}{Hg} cannot be created by the $s$ and $r$ process, which are responsible for the synthesis of the bulk of elements heavier than iron~\cite{1957Burbidge,1957Cameron}. The $s$ process, whose most probable astrophysical site are the AGB stars~\cite{KOLOCZEK20161}, runs close to the valley of stability, creating stable nuclei until Bismuth. On the contrary, the $r$ process rapidly deviates from the stable region, flowing to neutron-rich nuclei far from stability, in order to allow for subsequent $\beta^-$ decays to form neutron-rich nuclei. The kilonova signal~\cite{Smartt2017}, observed in 2017 by a neutron star merger has greatly contributed to the question of the astrophysical site of the $r$ process~\cite{Domoto_2021}. 

For the creation of the nuclei on the neutron-deficient side, another process had to be introduced in order to explain their existence in our solar system. This process is called the $p$ process, and when photodisintegration reactions are dominant, the term $\gamma$ process is also used~\cite{2003Arnould}. Candidates for the astrophysical sites of the $p$ process are the supernovae of type II and type Ia~\cite{1995Rayet,Travaglio2011, Choplin2022}.

The modeling of the reaction network of the $\gamma$ process consists of around 20,000 reactions involving approximately 2,000 nuclei below a mass number of $A \leq 210$~\cite{2003Arnould}.  Due to the scarcity of experimental cross sections of the photodisintegration reactions $(\gamma,n)$, $(\gamma,p)$, $(\gamma,\alpha)$, as well as their inverse ones $(n,\gamma)$, $(p,\gamma)$, $(\alpha,\gamma)$, the calculations of the relevant reaction rates rely heavily on Hauser-Feshbach statistical model calculations~\cite{Hauser_Feshbach_1952}. Within the Hauser-Feshbach model, quantities such as Optical Model Potentials (OMPs), Nuclear Level Densities (NLDs) and $\gamma$-strength functions ($\gamma$SFs) are parameters that can significantly influence the reaction cross sections and subsequently the reaction rates in the respective stellar environments.

At relatively low energies and close to the low-mass $p$ nuclei, the OMPs are the most sensitive parameter in cross section calculations~\cite{2006Rapp}. However, large uncertainties seem to arise in higher energies due to the unknown NLDs and $\gamma$SFs. These two parameters can be significantly constrained, when experimental data are available by the OSLO method~\cite{GUTTORMSEN1996371}. Data obtained by this method can help constrain NLD and $\gamma$SFs, which have a significant impact on the cross sections relevant to the $\gamma$-process reaction network.

In this work, we focus on the impact of NLD uncertainties on the radiative-capture cross sections for several important reactions for the $\gamma$ process~\cite{Rauscher2002,Rauscher2013}. In particular, the present work is concentrated on the radiative proton-capture reactions which can happen in the vicinity of the \textit{p} nuclei \isotope[102]{Pd} and \isotope[106]{Cd}. The reactions \isotope[104,105]{Ag}($p,\gamma$)\isotope[105,106]{Cd} and \isotope[104,105]{Rh}($p,\gamma$)\isotope[105,106]{Pd} are studied in the present work in terms of their proton-capture cross sections and their uncertainty after optimization of the corresponding NLD data. Cross section data for these reactions still remain unmeasured, as such data would be also hard to obtain, in particular with stable beams, as the target nuclei are not
stable. However, experimental data on NLDs exist for those nuclei~\cite{OSLO_data}, making it possible to constrain the model calculations for the radiative proton-capture cross section. 

Bayesian optimization has been used in the present work in order to estimate $95\%$ high density intervals on the level density data which are available at the OSLO database of experimental NLDs~\cite{OSLO_data}. The phenomenological Back-shifted Fermi Gas model (BSFG)~\cite{DILG1973269,GROSSJEAN1985113}, as well as the semi-microscopic level densities derived from Hartree-Fock-Bogoliubov plus the combinatorial (HFB+comb.) method~\cite{Goriely_2008_PhysRevC.78.064307} have been used for the estimation of these intervals. Then, these intervals have been used in order to calculate the corresponding cross sections inside the Gamow window~\cite{2007Iliadis}. The nuclear reaction code TALYS~\cite{2007Talys} has been used to calculate all values for the theoretical NLDs and cross sections presented in this work. 

It is important to note that only the uncertainties arising from the NLDs are considered in this work. Concerning the OMPs and $\gamma$SF models, the Koning-Delaroche potential~\cite{Koning2003} and the Kopecky-Uhl standard Lorentzian~\cite{1990Kopecky} have been used respectively, using the default values of their parameters in TALYS. Furthermore, this work does not consider the systematic uncertainties arising due to the model dependent normalisation of the level density data~\cite{Goriely_2022_PhysRevC.106.044315}. While the latter can give rise to important deviations, this work considers the data as given in the OSLO database~\cite{Egidy_bucurescu_2005_PhysRevC.72.044311}, along with the level density values at the neutron separation energy for each target nucleus given in~\cite{Eriksen_2014_PhysRevC.90.044311, Larsen_2013_PhysRevC.87.014319}. The normalization issue could be avoided by the newly developed shape method~\cite{Wiedeking_2021_PhysRevC.104.014311} in the future, which introduces a model-independent normalization of the level density data.

Radiative-proton capture reactions proceed mainly via the compound nucleus mechanism in the Gamow window. For the scenario of the synthesis of the $p$ nuclei during the explosive phase of a type II Supernova, where the temperature is in the range $T_9=1.7-3.3$~\cite{2003Arnould}, this window ranges from 1.2 to 6.4 MeV for the whole reaction network and in the center of mass for proton-capture reactions.In this range, the reactions proceed by the compound nucleus mechanism~\cite{Hodgson_1987,1990Satchler}. In Table~\ref{tab: reactions}, the Gamow Window for, as well as the Q-values and excitation energies for the compound nuclei formed after the reactions  \isotope[104,105]{Ag}($p,\gamma$)\isotope[105,106]{Cd} and \isotope[104,105]{Rh}($p,\gamma$)\isotope[105,106]{Pd} are given, along with the corresponding references of the NLD data.

\begin{table}[t]
\centering
\caption{Table of reactions relevant to the \textit{p} process studied in this work. The Gamow window (GW) corresponds to stellar temperatures of $T_9=1.7-3.3$.}
\begin{tabular}{ccccc}
\hline\noalign{\smallskip}
Reaction & Q-value & GW (c.m.)   & $E_x$ & Ref. (LD)  \\
\noalign{\smallskip}\hline\noalign{\smallskip}
\isotope[104]{Rh}($p,\gamma$)\isotope[105]{Pd} & 8.748 & 1.54-4.55 & 10.28-13.30 & \cite{Eriksen_2014_PhysRevC.90.044311} \\
\isotope[105]{Rh}($p,\gamma$)\isotope[106]{Pd} & 9.345 & 1.54-4.55 & 10.88-13.89 & \cite{Eriksen_2014_PhysRevC.90.044311}  \\
\isotope[104]{Ag}($p,\gamma$)\isotope[105]{Cd} & 6.506 & 1.59-4.67  & 8.10-11.17  & \cite{Larsen_2013_PhysRevC.87.014319} \\
\isotope[105]{Ag}($p,\gamma$)\isotope[106]{Cd} &  7.350 & 1.59-4.67 &  8.94-12.02 & \cite{Larsen_2013_PhysRevC.87.014319} \\

\noalign{\smallskip}\hline
\end{tabular}
\label{tab: reactions}
\end{table}

\section{Models and Methods}
\label{NLD models}

\subsection{The Back-shifted Fermi Gas Model}

\begin{table}[]
    \centering
    \caption{Prior distributions for parameters $a, \delta^{BSFG}$ of the BSFG model and $c,\delta^{HFB}$ of the scaled HFB+comb. model (Eq.~4). An uninformative gaussian prior is used in all cases. The means $a_M$, $\delta_M^{BSFG}, c_M, \delta_M^{HFB}$ and their standard deviations $\sigma_{a}, \sigma_{\delta^{BSFG}}, \sigma_c, \sigma_{\delta^{HFB}}$ are tabulated. See text for details.}
    \begin{tabular}{c|c|c|c|c|c|c|c|c|c|c}
        Isotope & $a_M$  & $\delta_M^{BSFG}$ & $\sigma_{a}$ & $\sigma_{\delta^{BSFG}} $ & $c_M$ & $\delta_M^{HFB}$ & $\sigma_c$ & $\sigma_{\delta^{HFB}}$ \\ \hline
        \isotope[105]{Pd} & 11.93 & 0.17 & 119.3 & 17 & 0 & 0 & 10 & 10 \\
        \isotope[106]{Pd} & 12.03 & 0.17 & 120.3 & 17 & 0 & 0 & 10 & 10 \\
        \isotope[105]{Cd} & 11.93 & 0.17 & 119.3 & 17 & 0 & 0 & 10 & 10 \\
        \isotope[106]{Cd} & 12.03 & 0.17 & 120.3 & 17 & 0 & 0 & 10 & 10
         
    \end{tabular}
   
    \label{tab: priors}
\end{table}

\begin{figure*}[t]
    \centering
    \begin{subfigure}{0.25\textwidth}
    \includegraphics[width=\textwidth]{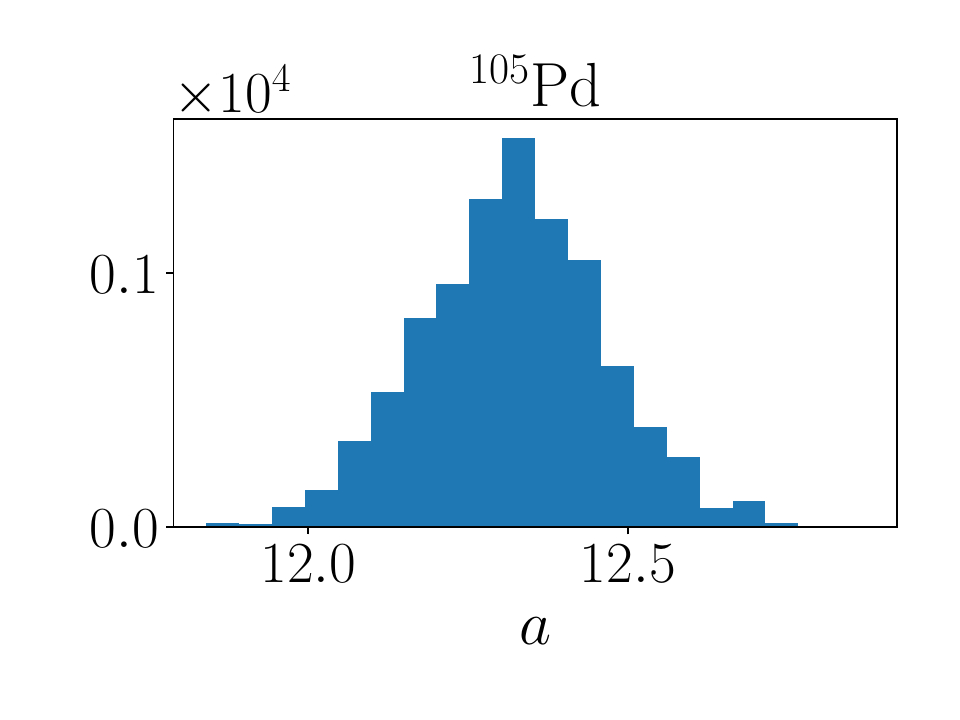}
    \caption{}
    \label{subfig: a ld 2 105Pd}
    \end{subfigure}%
      \begin{subfigure}{0.25\textwidth}
    \includegraphics[width=\textwidth]{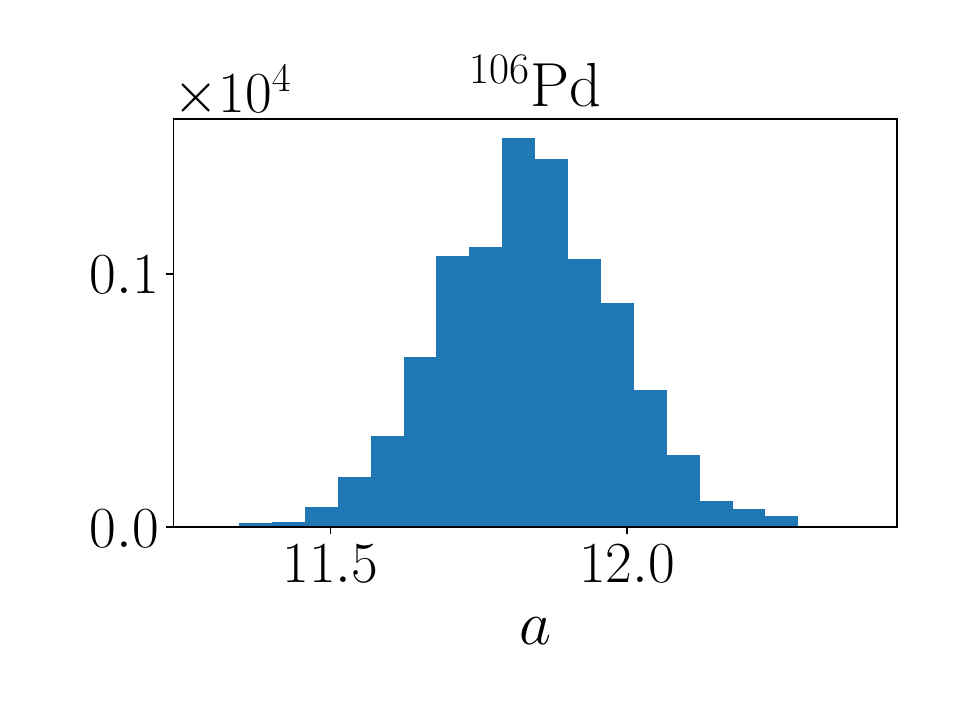}
    \caption{}
    \label{subfig: a ld2 106Pd}
    \end{subfigure}%
    \begin{subfigure}{0.25\textwidth}
    \includegraphics[width=\textwidth]{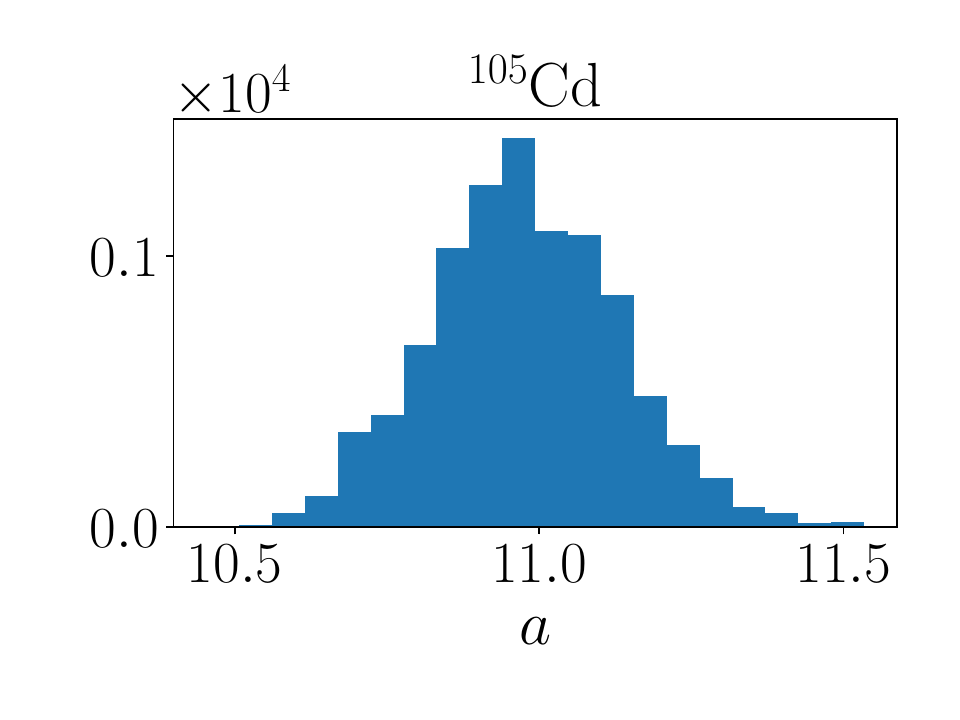}
    \caption{}
    \label{subfig: a ld2 105Cd}
    \end{subfigure}%
    \begin{subfigure}{0.25\textwidth}
    \includegraphics[width=\textwidth]{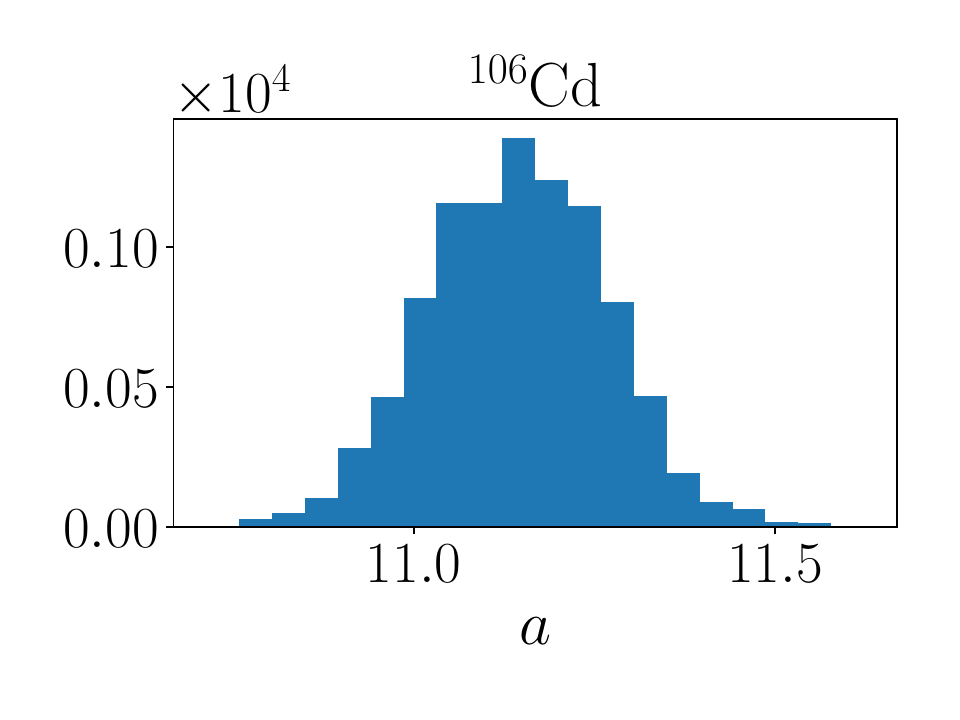}
    \caption{}
    \label{subfig: a ld2 106Cd}
    \end{subfigure}
      \begin{subfigure}{0.25\textwidth}
    \includegraphics[width=\textwidth]{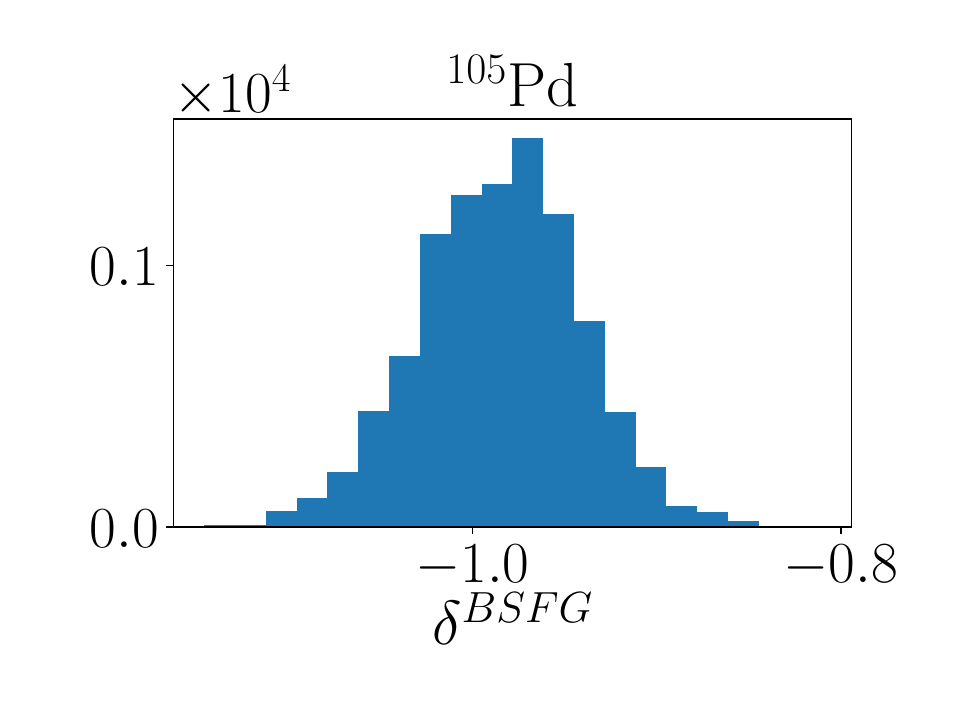}
    \caption{}
    \label{subfig: d ld2 105Pd}
    \end{subfigure}%
    \begin{subfigure}{0.25\textwidth}
    \includegraphics[width=\textwidth]{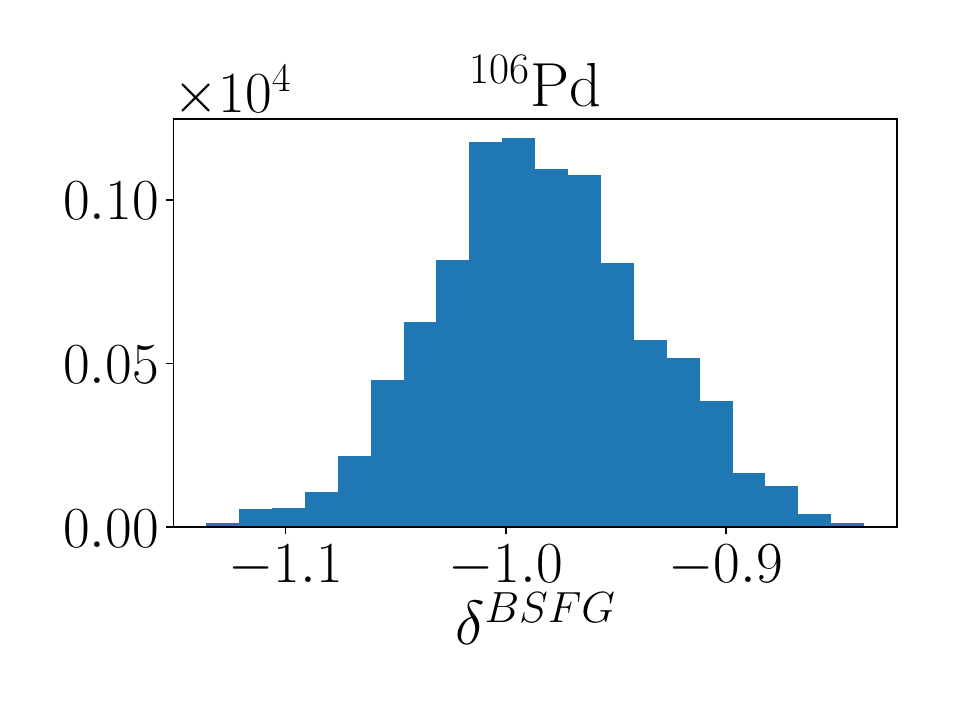}
    \caption{}
    \label{subfig: d ld2 106Pd}
    \end{subfigure}%
    \begin{subfigure}{0.25\textwidth}
    \includegraphics[width=\textwidth]{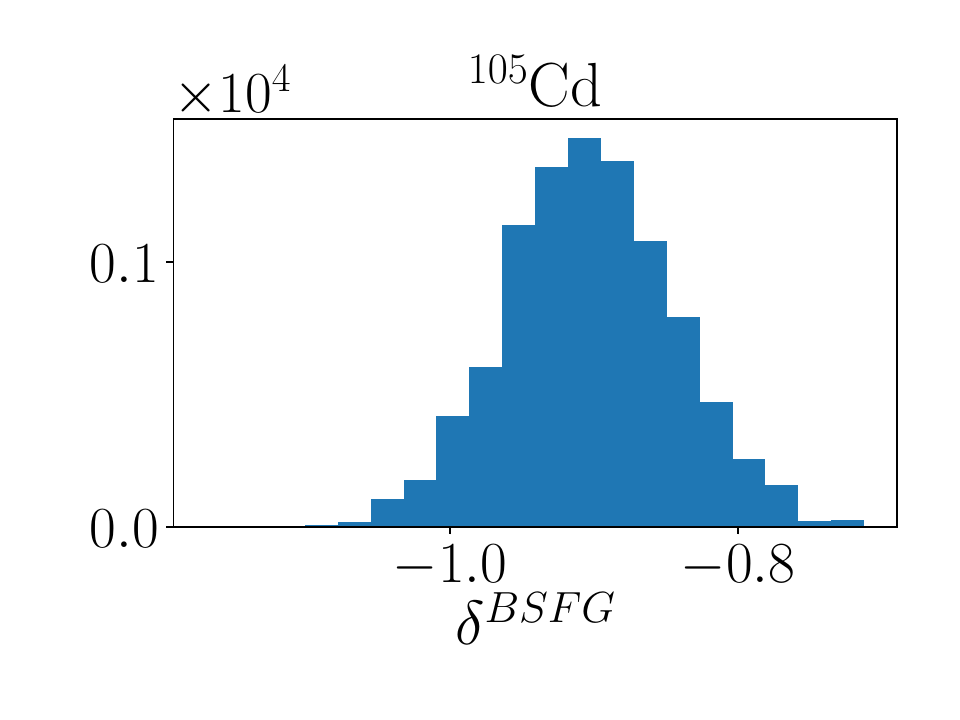}
    \caption{}
    \label{subfig: d ld2 105Cd}
    \end{subfigure}%
    \begin{subfigure}{0.25\textwidth}
    \includegraphics[width=\textwidth]{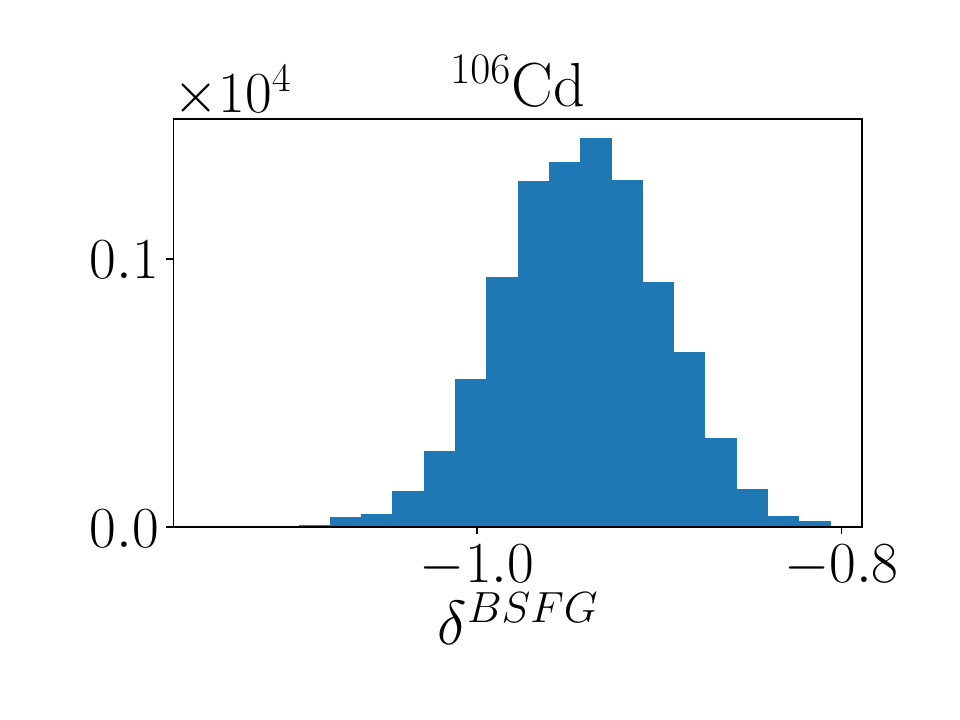}
    \caption{}
    \label{subfig: d ld2 106Cd}
    \end{subfigure}
    \begin{subfigure}{0.25\textwidth}
    \includegraphics[width=\textwidth]{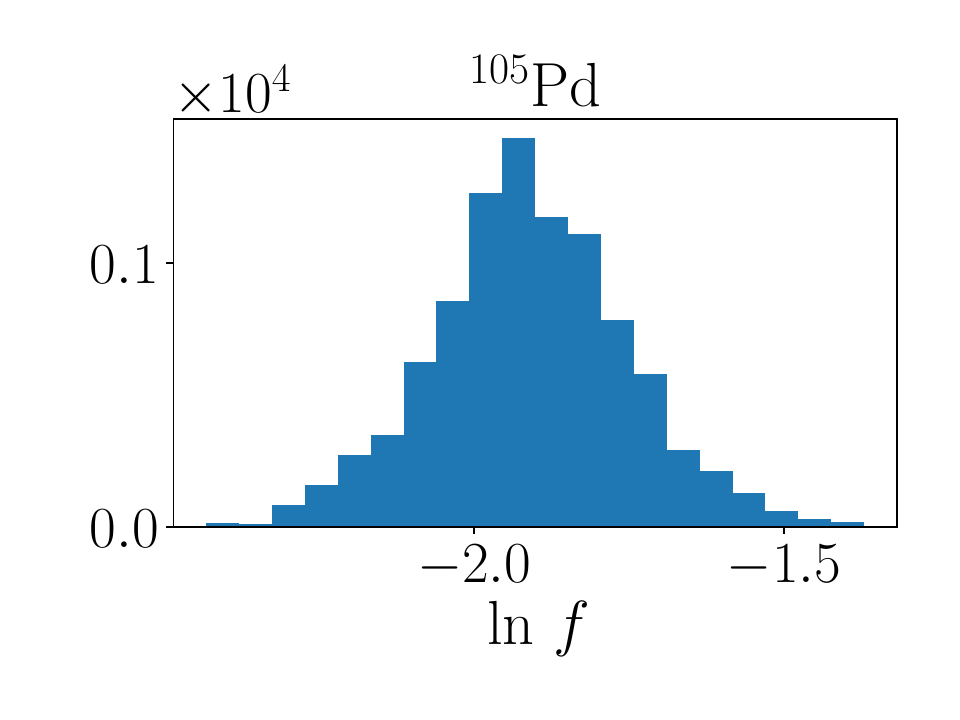}
    \caption{}
    \label{subfig: f ld2 105Pd}
    \end{subfigure}%
    \begin{subfigure}{0.25\textwidth}
    \includegraphics[width=\textwidth]{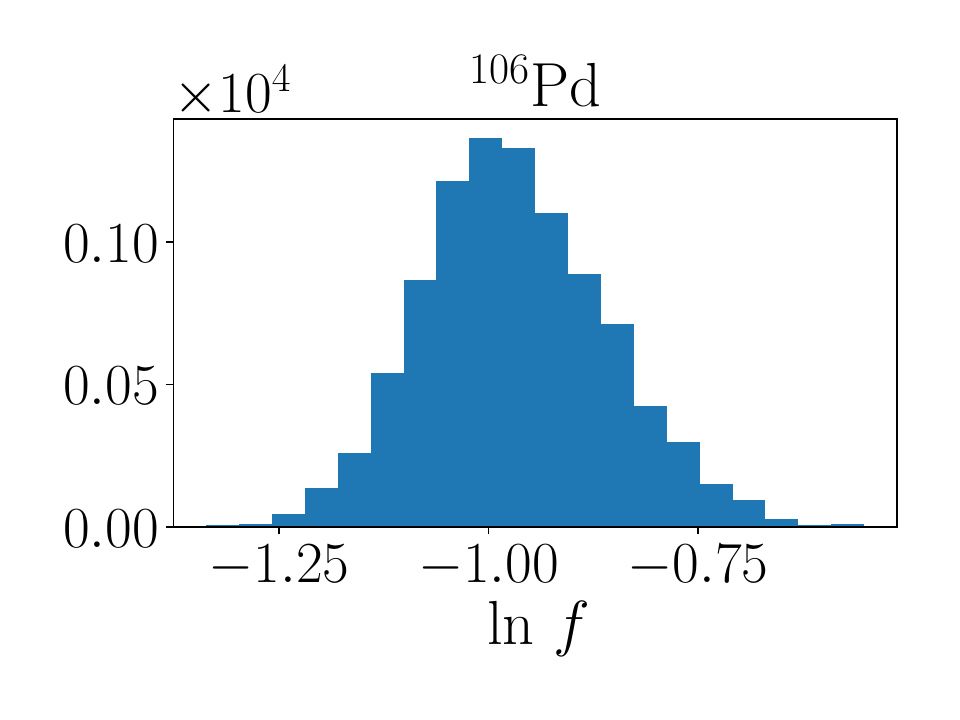}
    \caption{}
    \label{subfig: f ld2 106Pd}
    \end{subfigure}%
    \begin{subfigure}{0.25\textwidth}
    \includegraphics[width=\textwidth]{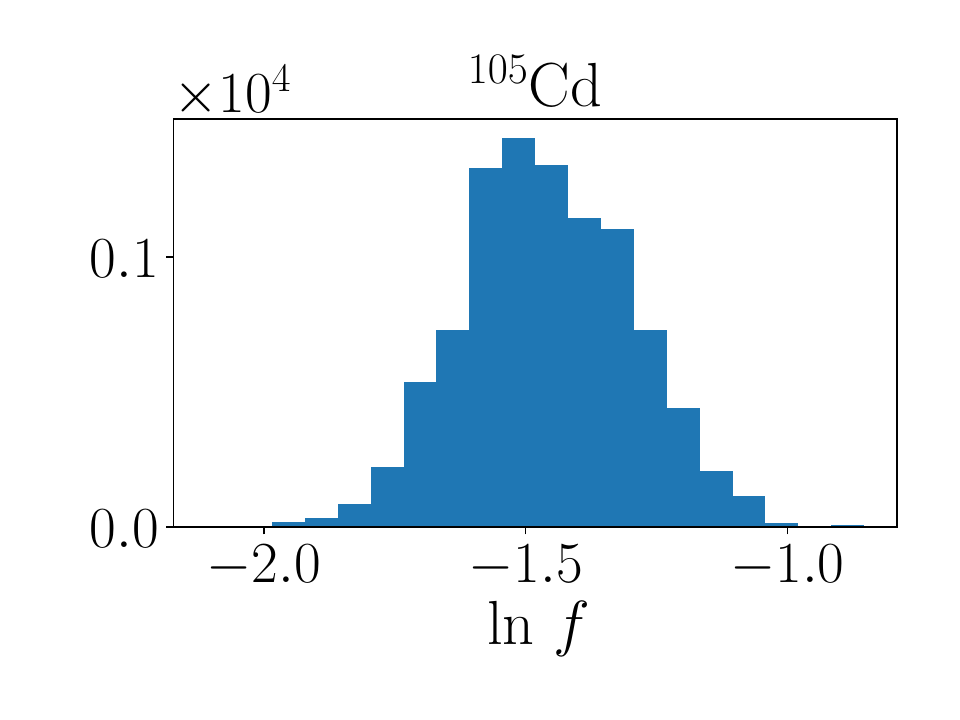}
    \caption{}
    \label{subfig: f ld2 105Cd}
    \end{subfigure}%
    \begin{subfigure}{0.25\textwidth}
    \includegraphics[width=\textwidth]{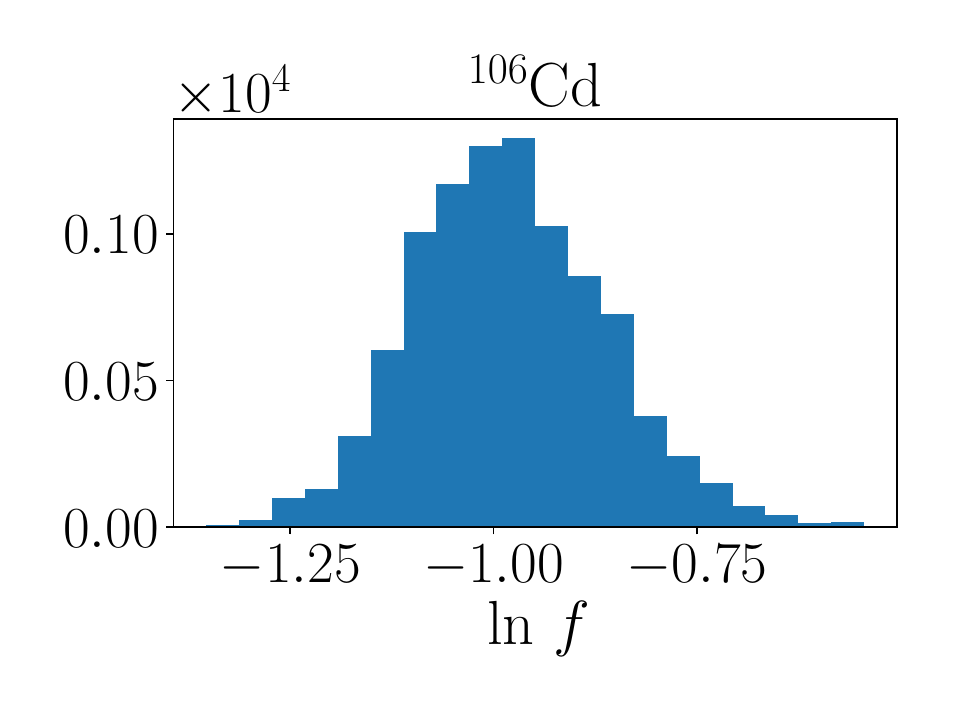}
    \caption{}
    \label{subfig: f ld2 106Cd}
    \end{subfigure}%

    \caption{Posterior distributions of the parameters $a$, $\delta^{BSFG}$ and $\ln f$ by optimizing the BSFG model on the OSLO data for the isotopes \isotope[105, 106]{}{Pd} and \isotope[105, 106]{}{Cd}. See text for details.}
    \label{fig: posteriors bsfg}
\end{figure*}

The first model that will be used in the present work in order to constrain the level densities is the phenomenological Back-shifted Fermi Gas model (BSFG)~\cite{DILG1973269}. The advantage of this phenomenological model allows for the optimization of its parameters along the whole energy range, contrary to the Gilbert-Cameron model~\cite{Gilbert1965}, which is essentially a combination of the Constant Temperature model in low excitation energies and the Fermi Gas model in higher excitation energies. Furthermore, the OSLO data used in this work have been renormalized using this model~\cite{Egidy_bucurescu_2005_PhysRevC.72.044311}. In addition, the relevant excitation energies for proton capture relevant to the \textit{p} process are in the range where a Fermi gas description is more appropriate.

In the BSFG model, which is implemented in the TALYS code, the level density is given by the formula:
\begin{equation}\label{eq: ld bsfg}
    \rho_{tot} (U) = \frac{1}{\sqrt{2 \pi} \sigma} \frac{\sqrt{\pi}}{12} \frac{ \exp{2 \sqrt{aU}} }{a^{1/4}U^{5/4}}.
\end{equation}
where $\sigma$ is the spin-cutoff parameter. The quantity $U$ is connected with the excitation energy $E_x$:
\begin{equation}\label{eq: excitation energy}
U = E_x - \chi \frac{12}{\sqrt{A} + \delta^{BSFG}}
\end{equation}
where $\chi=-1,0,+1$ for odd-odd, odd-even and even even nuclei respectively and $a, \delta^{BSFG}$ are free parameters. The posterior distributions of these parameters will be sampled in order to estimate the high density intervals for each parameter. It is to be noted that while the parameter $a$ can be treated as energy dependent to account for shell effects, no energy dependence is taken into account in the present work.

\subsection{The Skyrme-Hartree-Fock Bogoliubov plus combinatorial Model}

The Skyrme-Hartree-Fock Bogoliubov plus combinatorial Model (HFB+comb.) is a microscopic model used to calculate the level densities for excitation energies up to 200 MeV~\cite{Goriely_2008_PhysRevC.78.064307}. These level densities are available in TALYS in tabulated format.

Within TALYS, the possibility of adjusting these microscopic level densities $\rho_{tab}$ to the experimental data has also been added, by using two-parameter scaling function of the form~\cite{2007Talys}:
\begin{equation}\label{eq: scaling function}
    \rho(E_x,J,\pi) = \exp(c \sqrt{E_x-\delta^{HFB}} \rho_{tab}(E_x-\delta^{HFB},J,\pi)
\end{equation}
where $c,\delta^{HFB}$ are parameters that play a similar role as the parameters $a$ and $\delta^{BSFG}$ of the previously mentioned BSFG model. 

\begin{figure*}[t]
    \centering
    \begin{subfigure}{0.25\textwidth}
    \includegraphics[width=\textwidth]{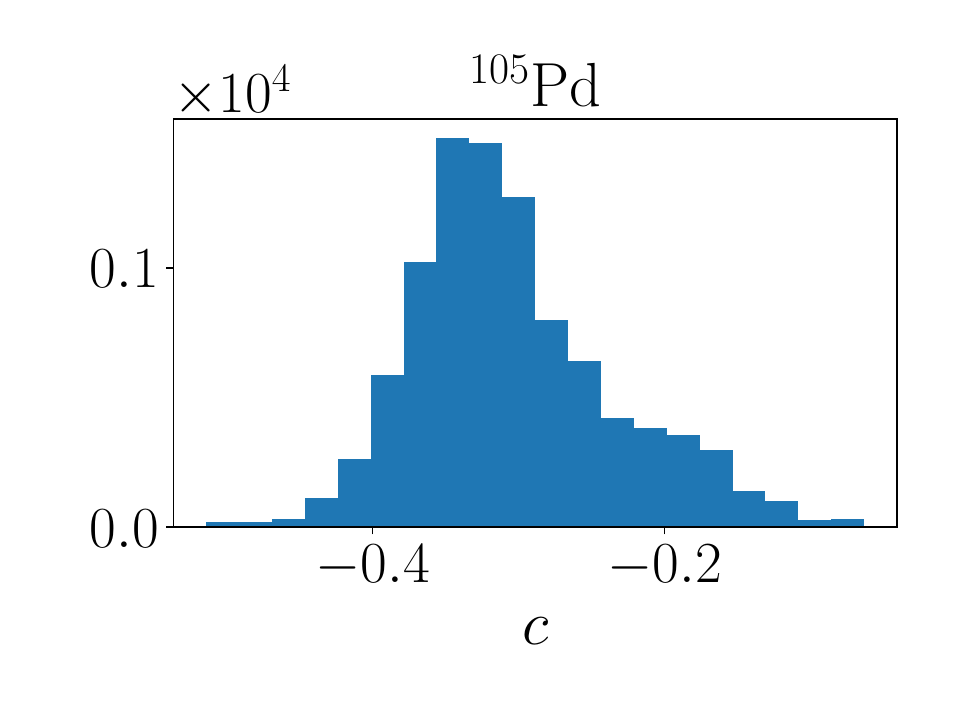}
    \caption{}
    \label{subfig: c ld 5 105Pd}
    \end{subfigure}%
      \begin{subfigure}{0.25\textwidth}
    \includegraphics[width=\textwidth]{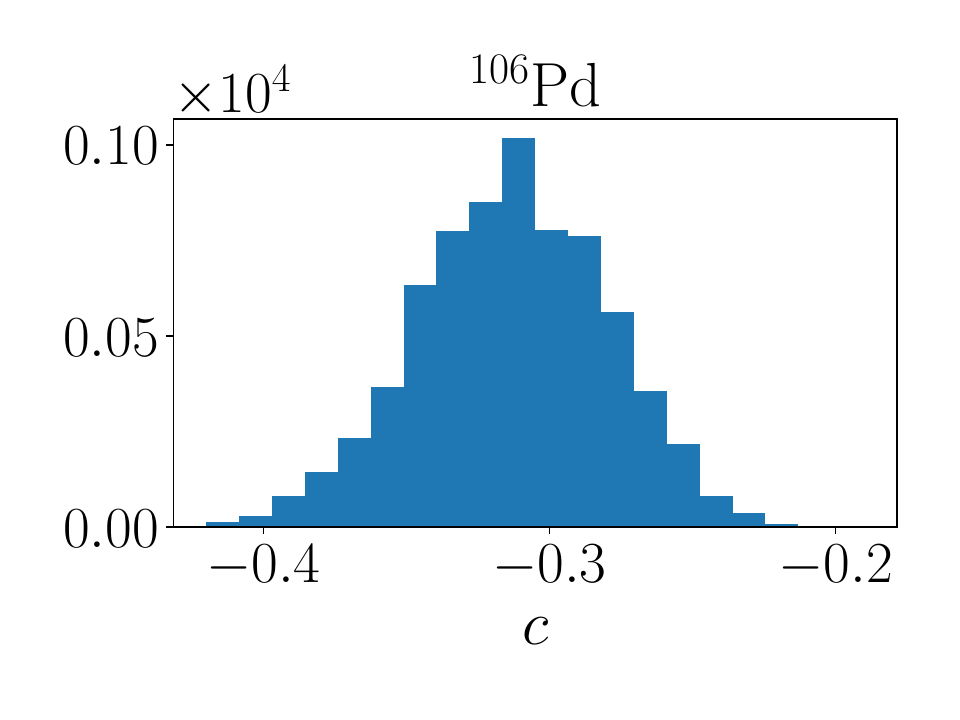}
    \caption{}
    \label{subfig: c ld5 106Pd}
    \end{subfigure}%
    \begin{subfigure}{0.25\textwidth}
    \includegraphics[width=\textwidth]{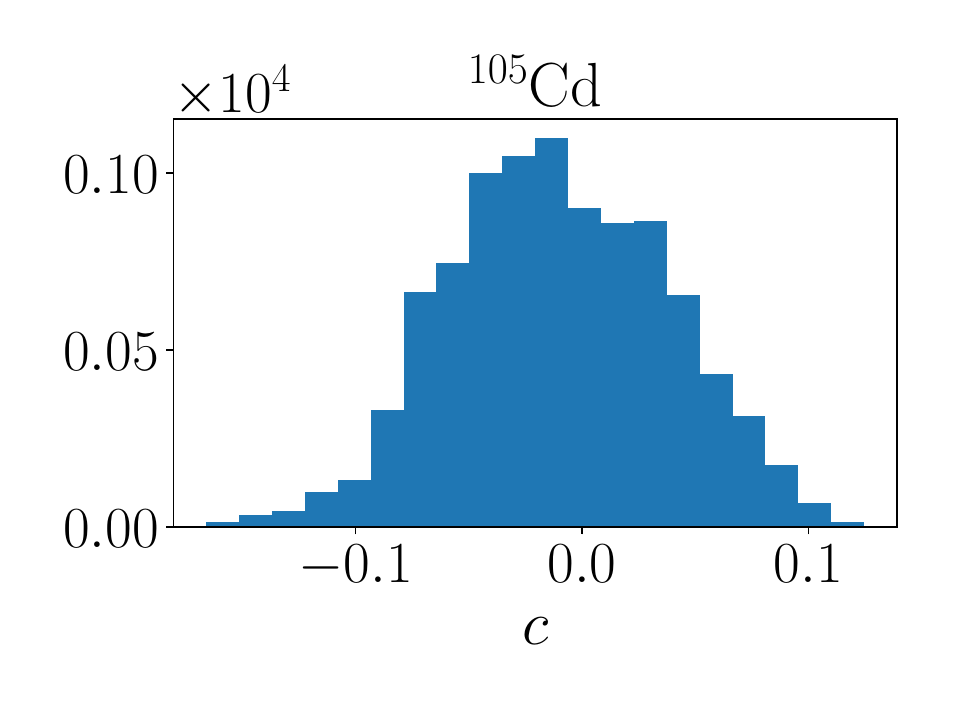}
    \caption{}
    \label{subfig: c ld5 105Cd}
    \end{subfigure}%
    \begin{subfigure}{0.25\textwidth}
    \includegraphics[width=\textwidth]{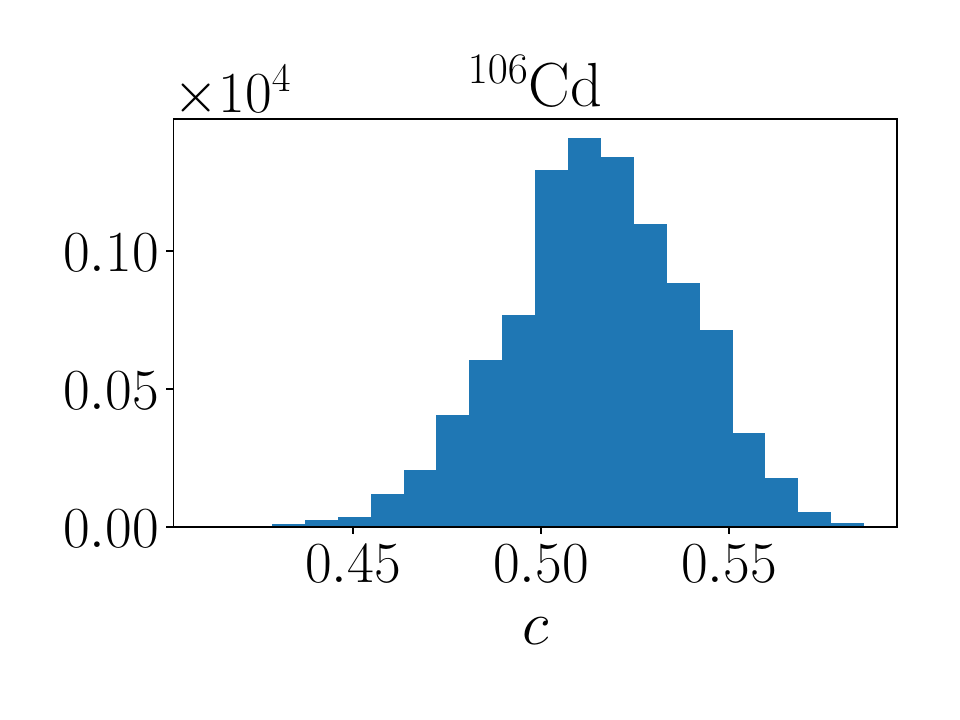}
    \caption{}
    \label{subfig: c ld5 106Cd}
    \end{subfigure}
      \begin{subfigure}{0.25\textwidth}
    \includegraphics[width=\textwidth]{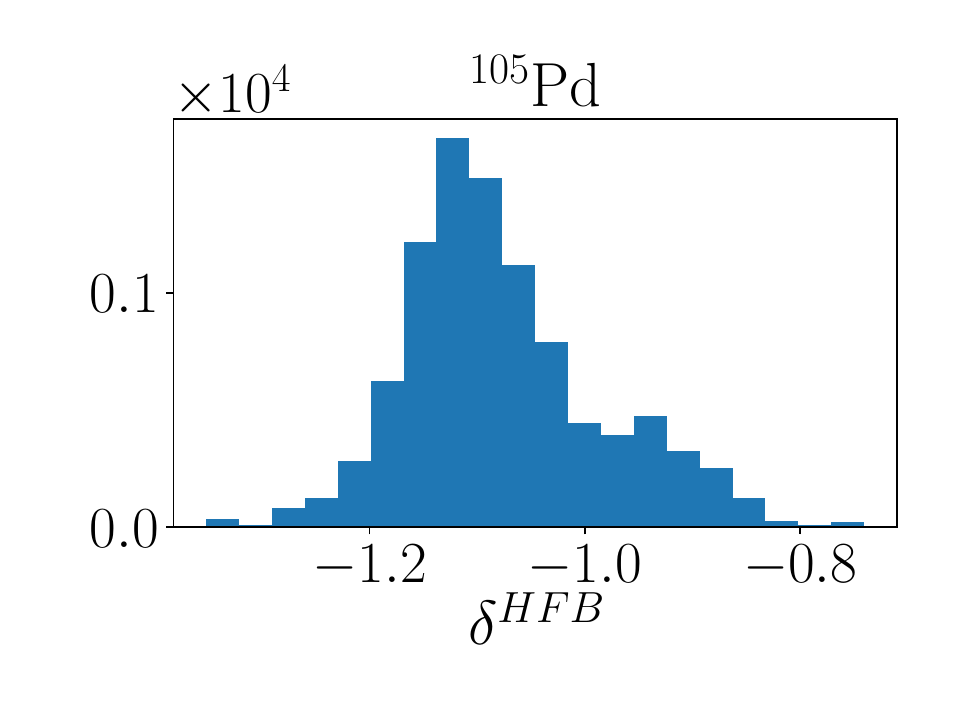}
    \caption{}
    \label{subfig: d ld5 105Pd}
    \end{subfigure}%
    \begin{subfigure}{0.25\textwidth}
    \includegraphics[width=\textwidth]{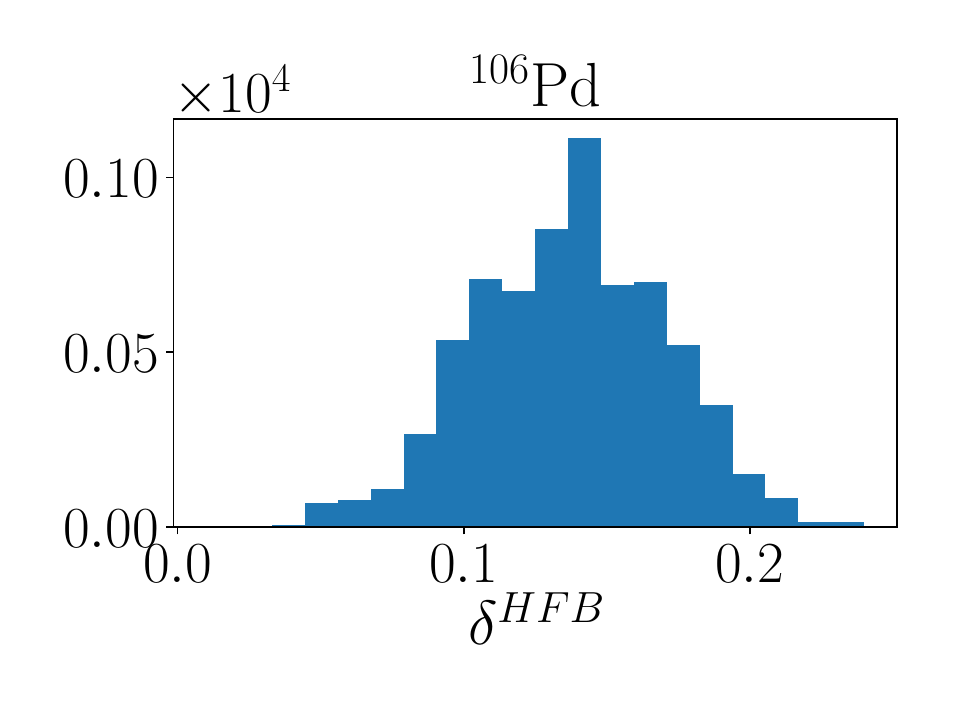}
    \caption{}
    \label{subfig: d ld5 106Pd}
    \end{subfigure}%
    \begin{subfigure}{0.25\textwidth}
    \includegraphics[width=\textwidth]{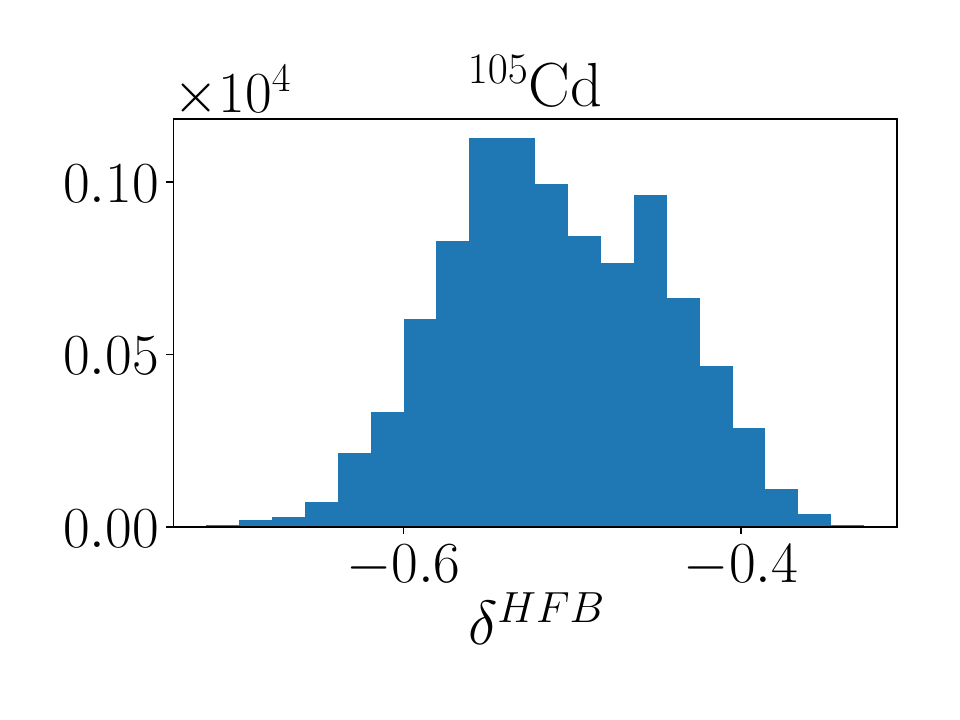}
    \caption{}
    \label{subfig: d ld5 105Cd}
    \end{subfigure}%
    \begin{subfigure}{0.25\textwidth}
    \includegraphics[width=\textwidth]{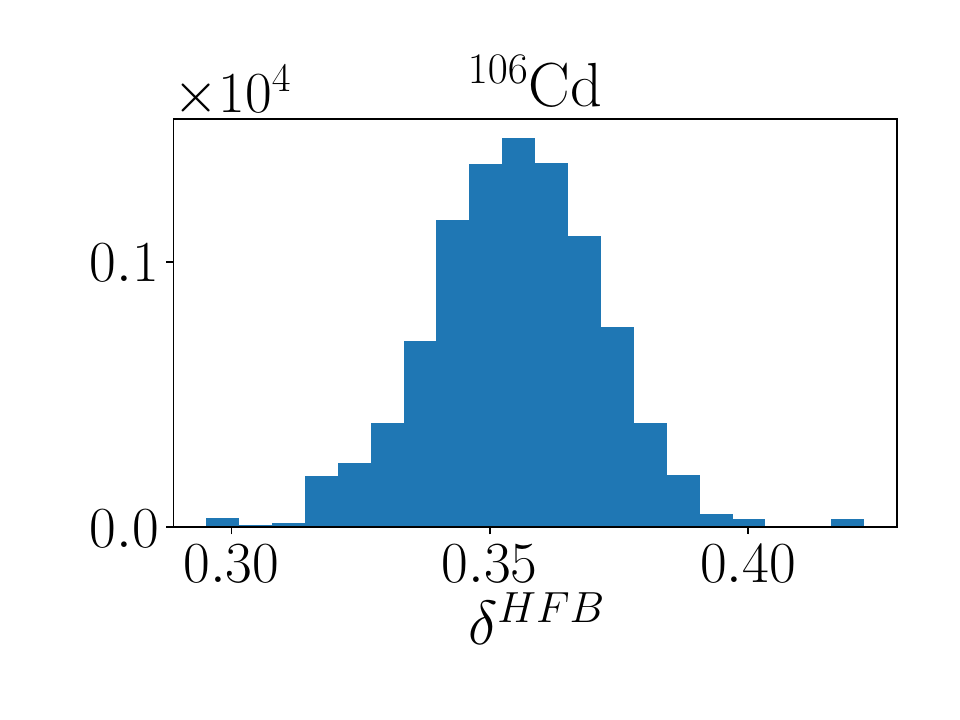}
    \caption{}
    \label{subfig: d ld5 106Cd}
    \end{subfigure}
       \begin{subfigure}{0.25\textwidth}
    \includegraphics[width=\textwidth]{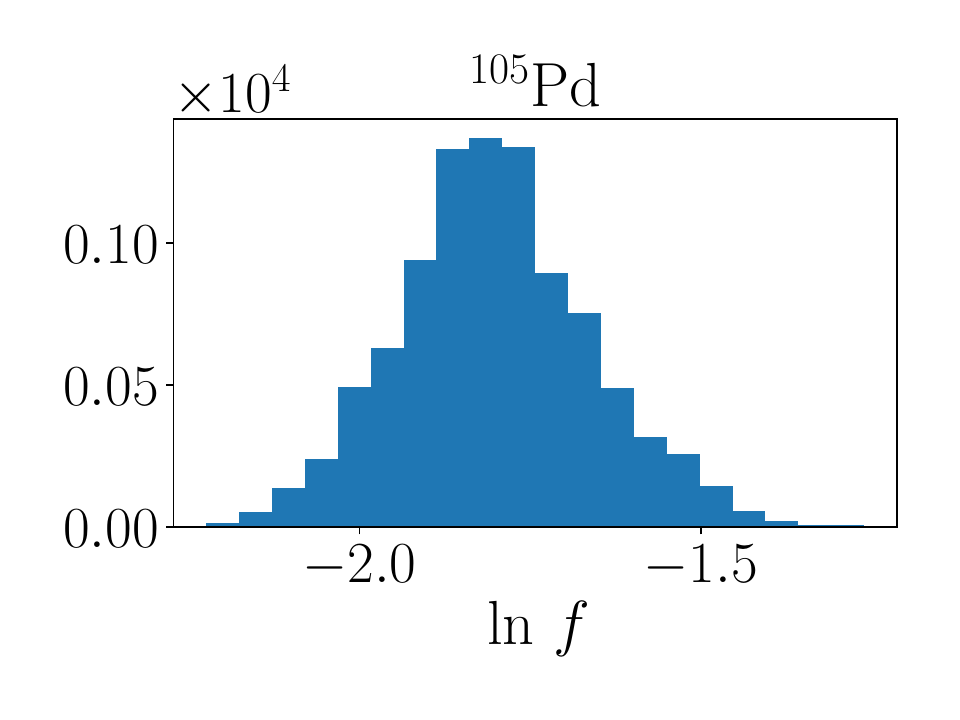}
    \caption{}
    \label{subfig: f ld5 105Pd}
    \end{subfigure}%
    \begin{subfigure}{0.25\textwidth}
    \includegraphics[width=\textwidth]{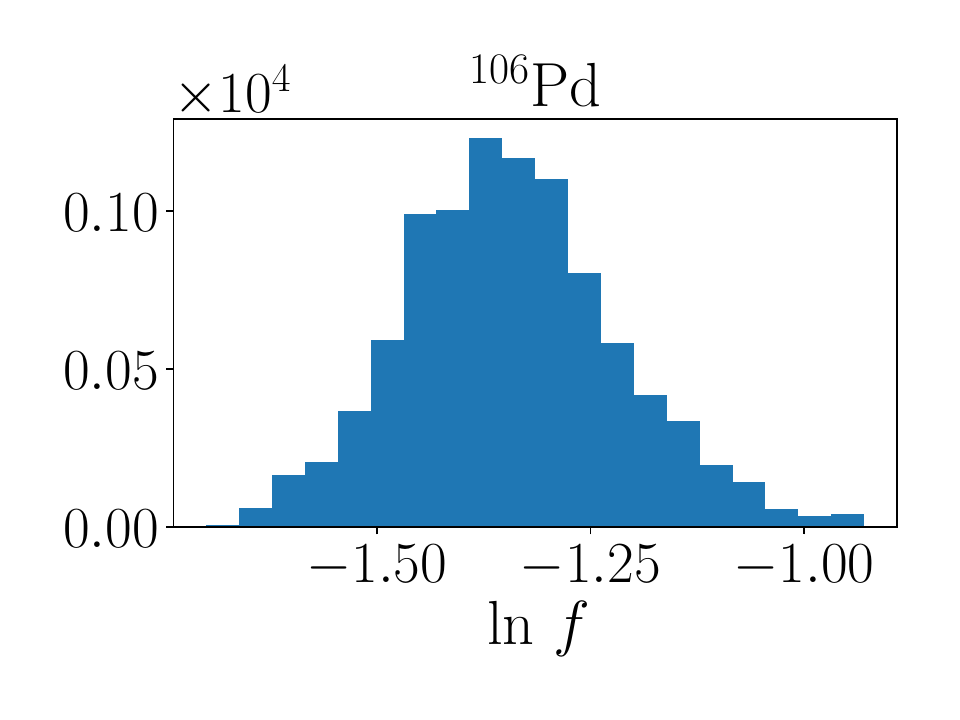}
    \caption{}
    \label{subfig: f ld5 106Pd}
    \end{subfigure}%
    \begin{subfigure}{0.25\textwidth}
    \includegraphics[width=\textwidth]{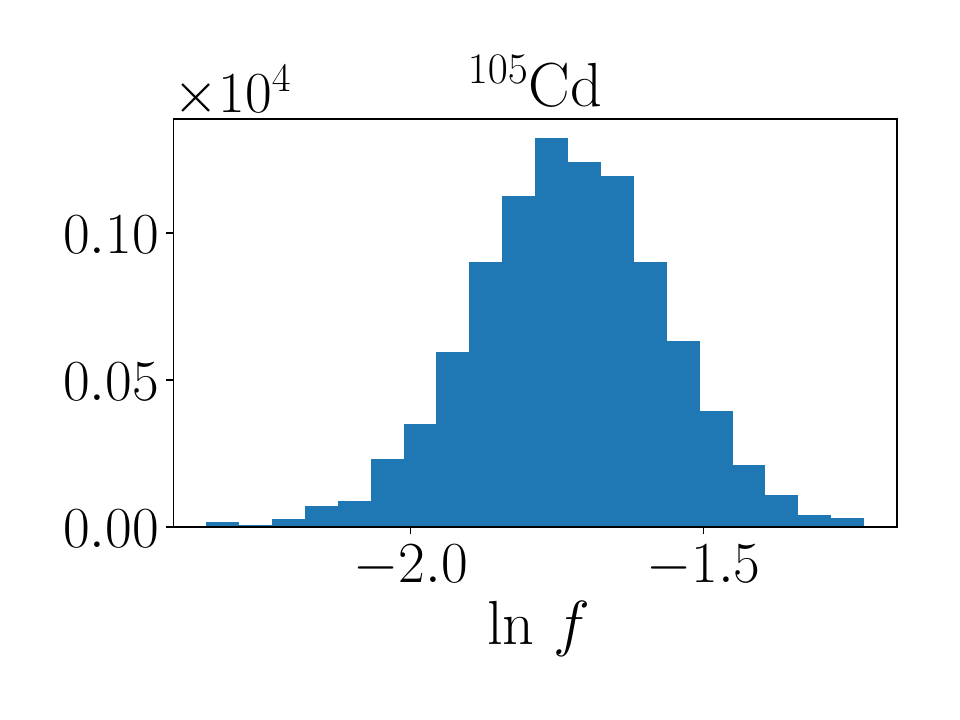}
    \caption{}
    \label{subfig: f ld5 105Cd}
    \end{subfigure}%
    \begin{subfigure}{0.25\textwidth}
    \includegraphics[width=\textwidth]{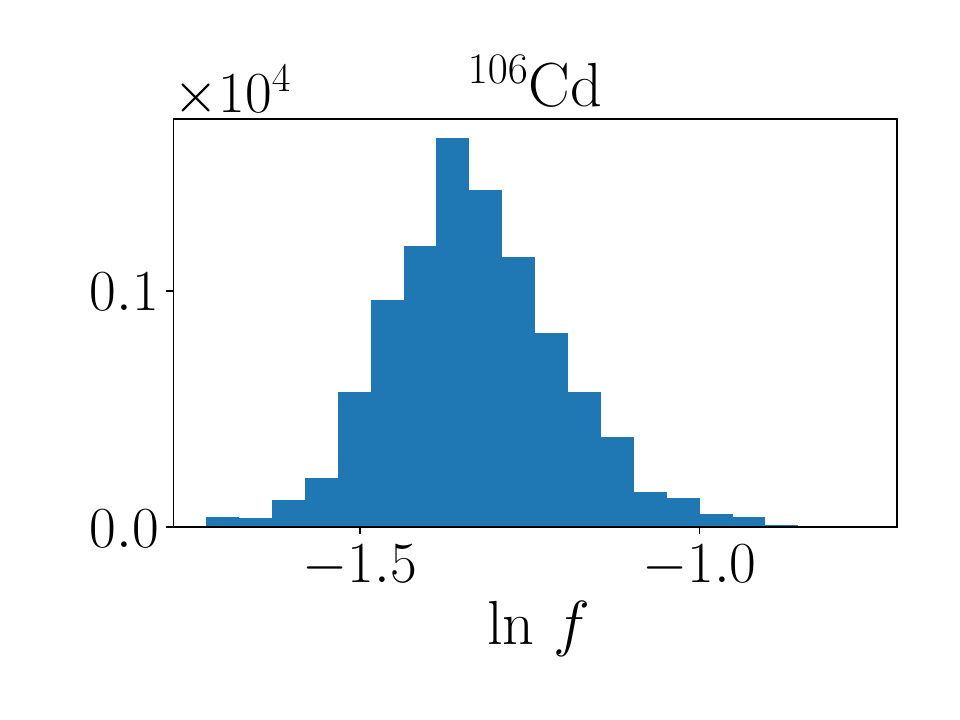}
    \caption{}
    \label{subfig: f ld5 106Cd}
    \end{subfigure}%

    \caption{Posterior distributions of the parameters $c$, $\delta^{HFB}$ and $\ln f$ by optimizing the scaling function of Eq.~\ref{eq: scaling function} for the HFB+comb. model on the OSLO data for the isotopes \isotope[105, 106]{}{Pd} and \isotope[105, 106]{}{Cd}. See text for details.}
    \label{fig: posteriors hfb}
\end{figure*}

Using the available experimental data given by the OSLO database, it is thus possible to calculate the posterior distributions for the parameters $c,\delta^{HFB}$ of the scaling function and estimate the relevant high density intervals. A comparison of the relevant high density intervals with the ones of the BSFG model is also interesting in order to see how the uncertainties vary depending on the model choice.

\begin{figure*}[t]
    \centering
    \begin{subfigure}{0.50\textwidth}
    \includegraphics[width=\textwidth]{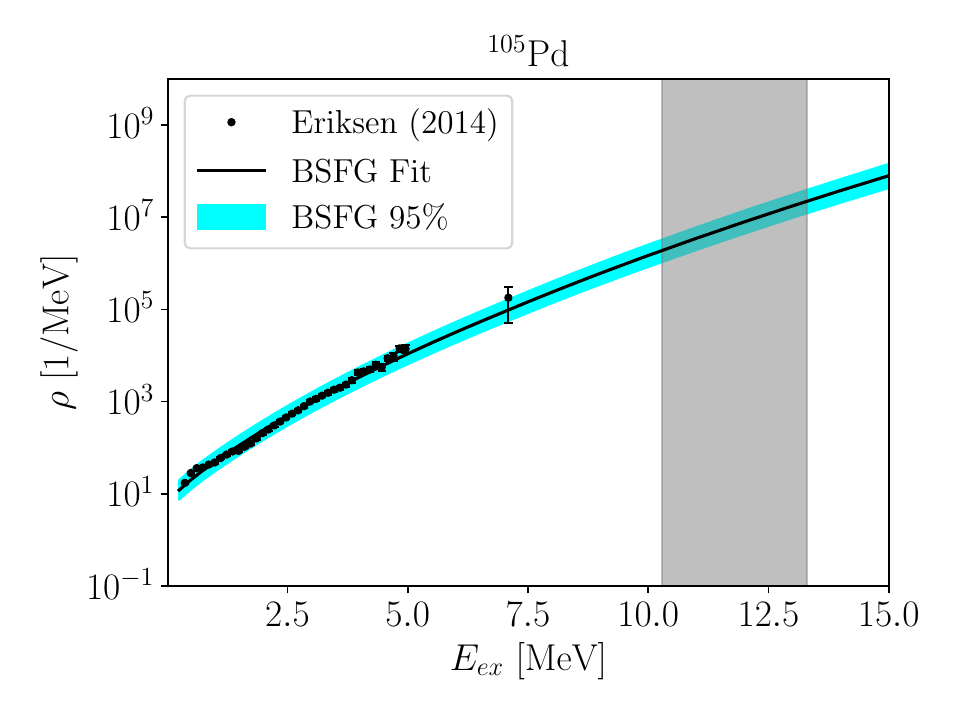}
    \caption{}
    \label{subfig: ld bsfg 105Pd}
    \end{subfigure}%
      \begin{subfigure}{0.50\textwidth}
    \includegraphics[width=\textwidth]{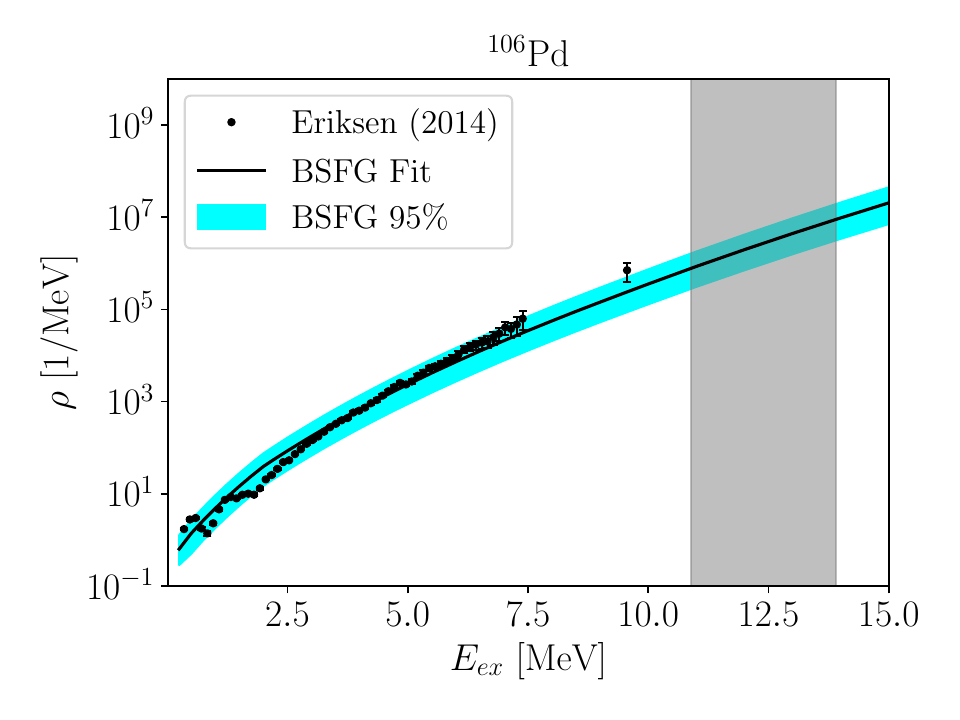}
    \caption{}
    \label{subfig: ld bsfg 106Pd}
    \end{subfigure}
    \begin{subfigure}{0.50\textwidth}
    \includegraphics[width=\textwidth]{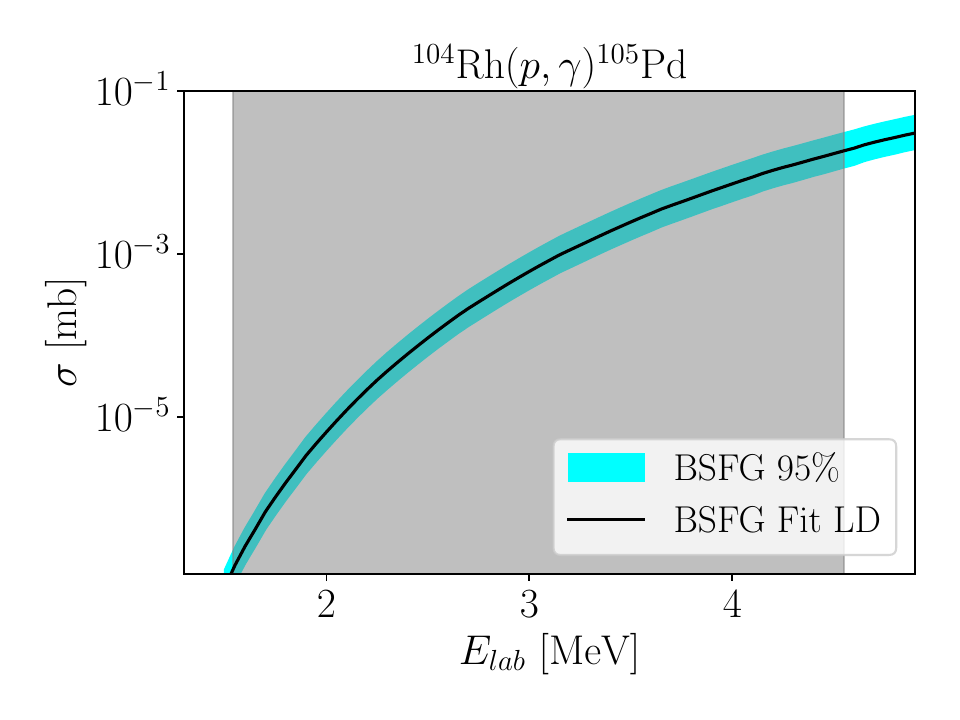}
    \caption{}
    \label{subfig: pg bsfg 105Pd}
    \end{subfigure}%
    \begin{subfigure}{0.5\textwidth}
    \includegraphics[width=\textwidth]{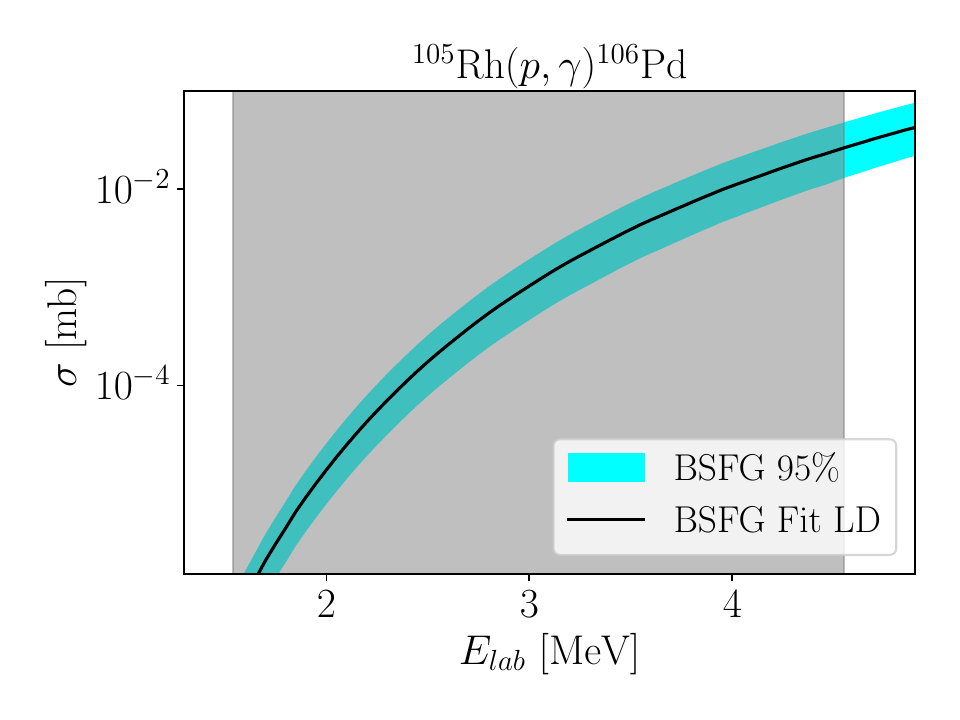}
    \caption{}
    \label{subfig: pg bsfg 106Pd}
    \end{subfigure}
     
    \caption{Calculated level densities and cross sections using the posterior distributions of the parameters $a,\delta^{BSFG}$ and $f$ of the BSFG model for the isotopes \isotope[105, 106]{}{Pd}. The best-fit value and the 95\% high density intervals obtained after Bayesian optimisation are shown with the black curve and the cyan band, respectively. The grey-shaded area represents the Gamow window for the reactions \isotope[104,105]{Rh}($p,\gamma$)\isotope[105,106]{Pd}. See text for details.}
    \label{fig: pd bsfg}
\end{figure*}

\begin{figure*}[t]
    \centering
    \begin{subfigure}{0.5\textwidth}
    \includegraphics[width=\textwidth]{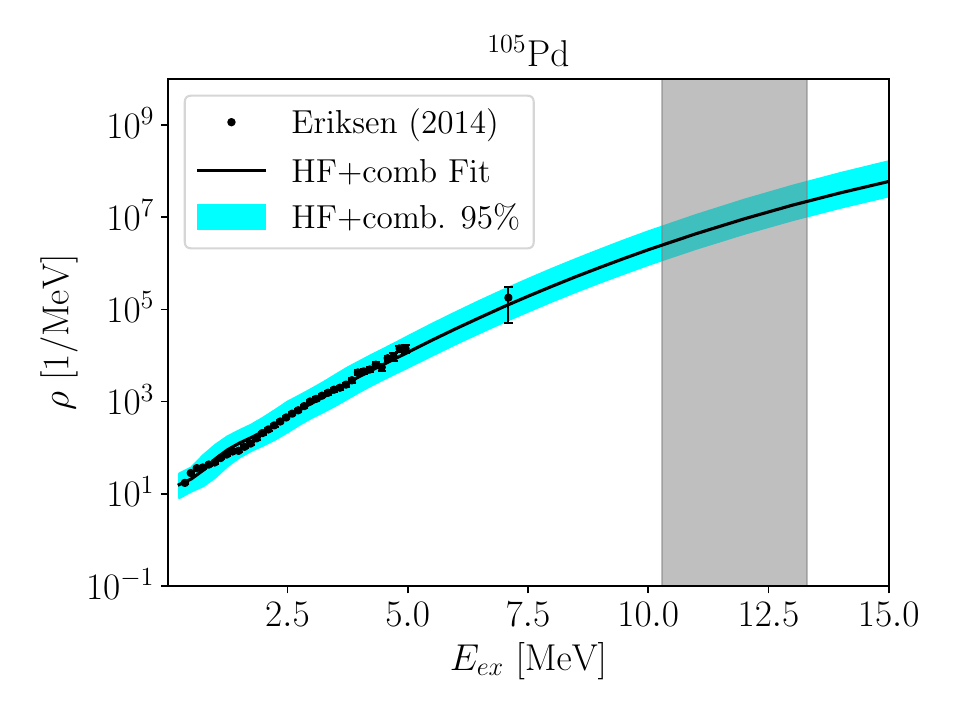}
    \caption{}
    \label{subfig: ld hfb 105Pd}
    \end{subfigure}%
    \begin{subfigure}{0.5\textwidth}
    \includegraphics[width=\textwidth]{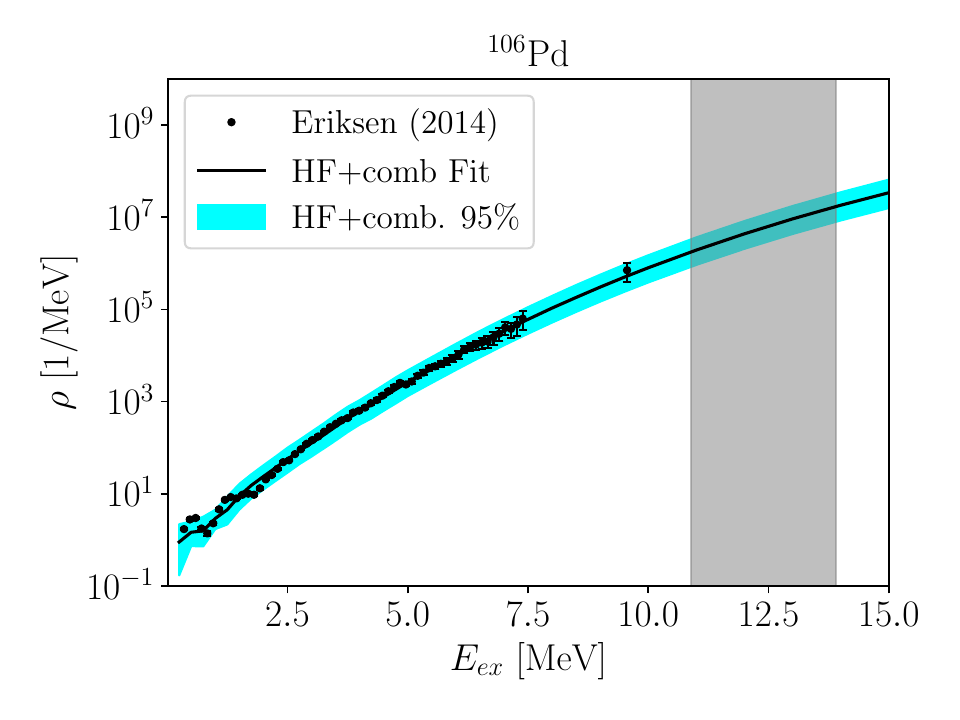}
    \caption{}
    \label{subfig: ld hfb 106Pd}
    \end{subfigure}
    \begin{subfigure}{0.5\textwidth}
    \includegraphics[width=\textwidth]{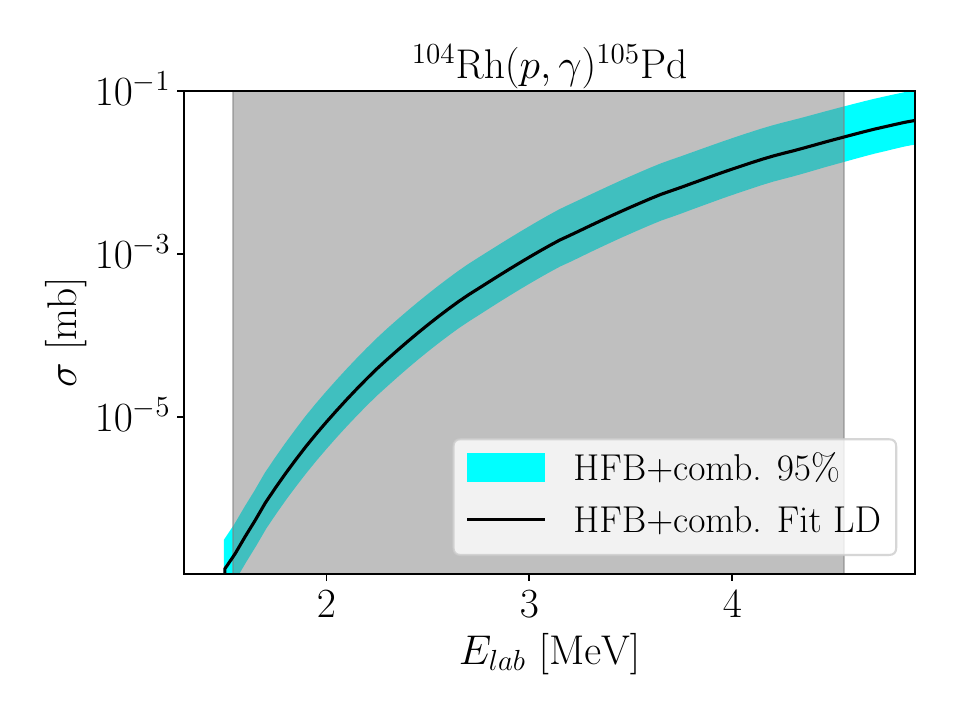}
    \caption{}
    \label{subfig: pg hfb 105Pd}
    \end{subfigure}%
    \begin{subfigure}{0.5\textwidth}
    \includegraphics[width=\textwidth]{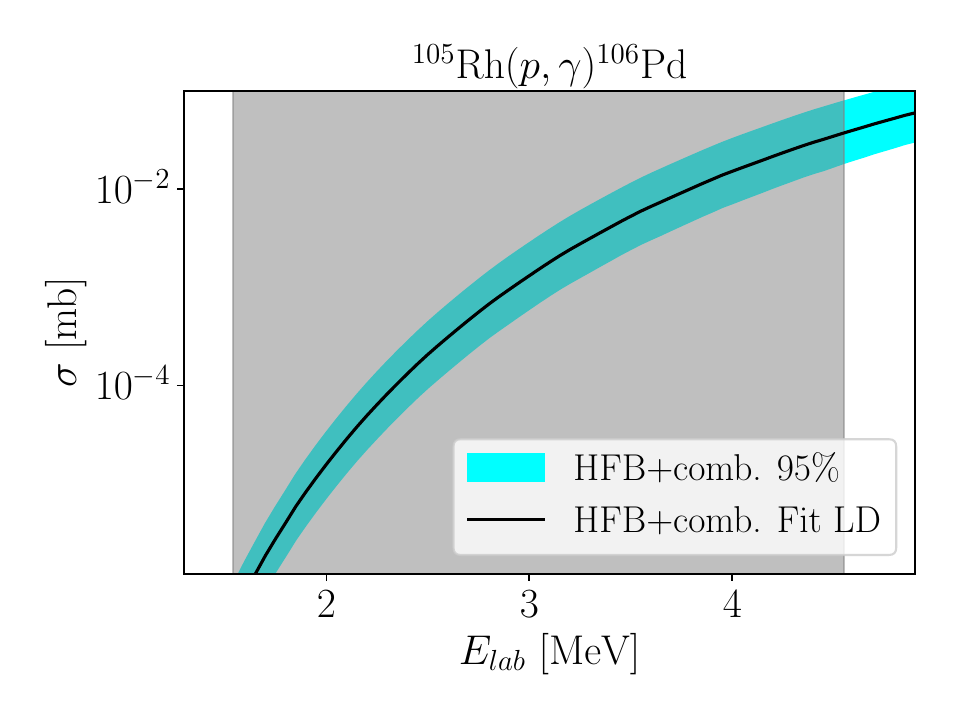}
    \caption{}
    \label{subfig: pg hfb 106Pd}
    \end{subfigure}    
    \caption{Calculated level densities and cross sections using the posterior distributions of the parameters $a,\delta^{HFB}$ and $f$ of the HFB+comb. model for the isotopes \isotope[105, 106]{}{Pd}. The best-fit value and the 95\% high density intervals obtained after Bayesian optimisation are shown with the black curve and the cyan band, respectively. The grey-shaded area represents the Gamow window for the reactions \isotope[104,105]{Rh}($p,\gamma$)\isotope[105,106]{Pd}. See text for details.} 
    \label{fig: pd hfb}
\end{figure*}

\begin{figure*}[t]
    \centering
    \begin{subfigure}{0.50\textwidth}
    \includegraphics[width=\textwidth]{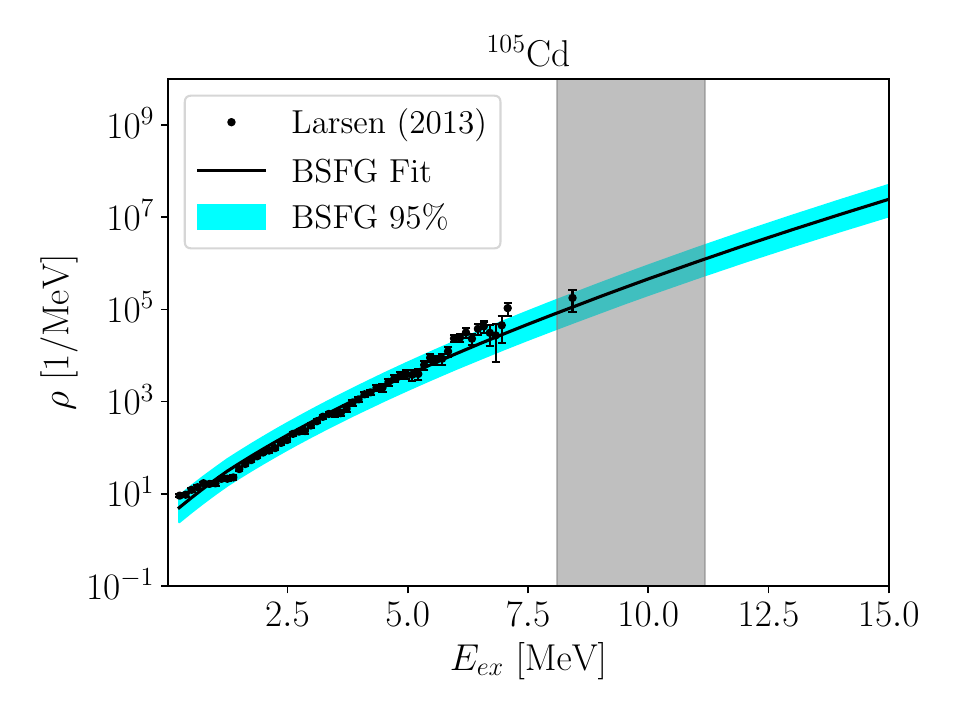}
    \caption{}
    \label{subfig: ld bsfg 105Cd}
    \end{subfigure}%
      \begin{subfigure}{0.50\textwidth}
    \includegraphics[width=\textwidth]{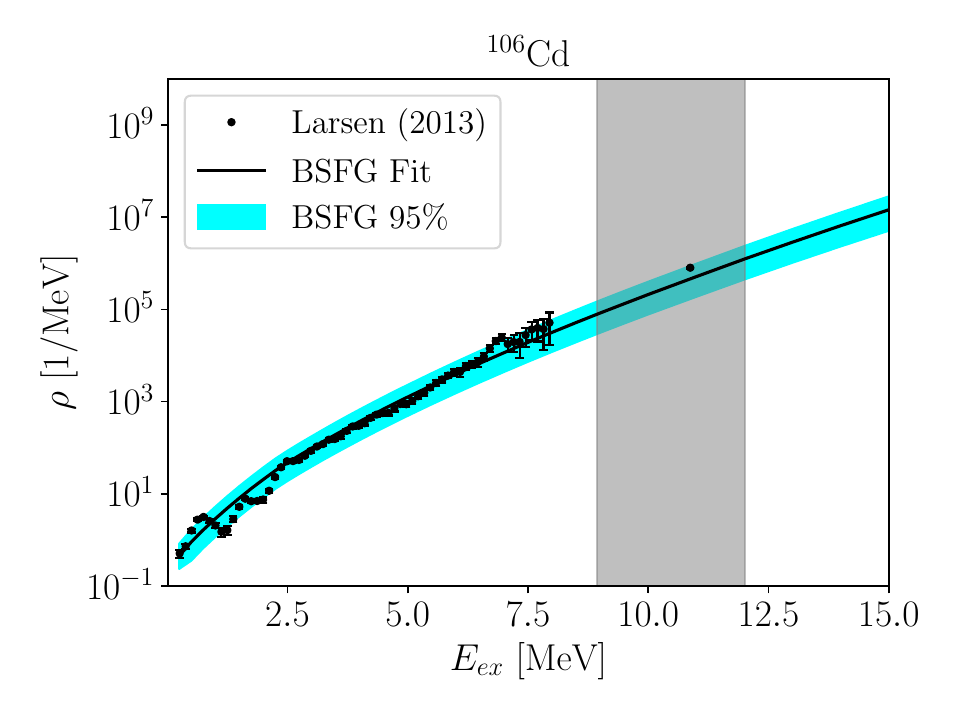}
    \caption{}
    \label{subfig: ld bsfg 106Cd}
    \end{subfigure}
    \begin{subfigure}{0.50\textwidth}
    \includegraphics[width=\textwidth]{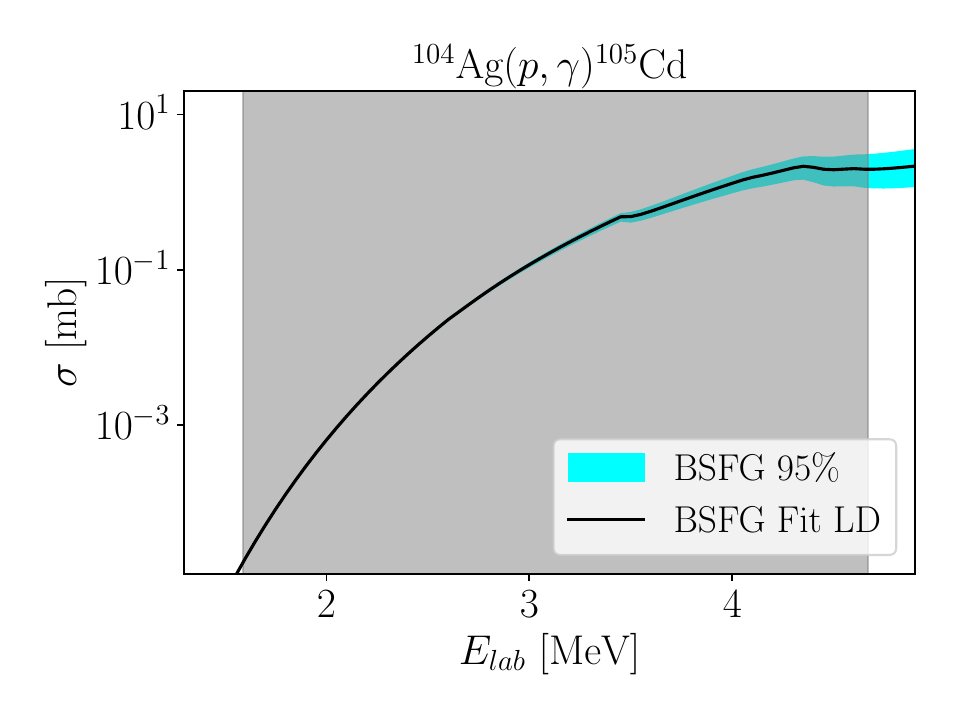}
    \caption{}
    \label{subfig: pg bsfg 105Cd}
    \end{subfigure}%
    \begin{subfigure}{0.5\textwidth}
    \includegraphics[width=\textwidth]{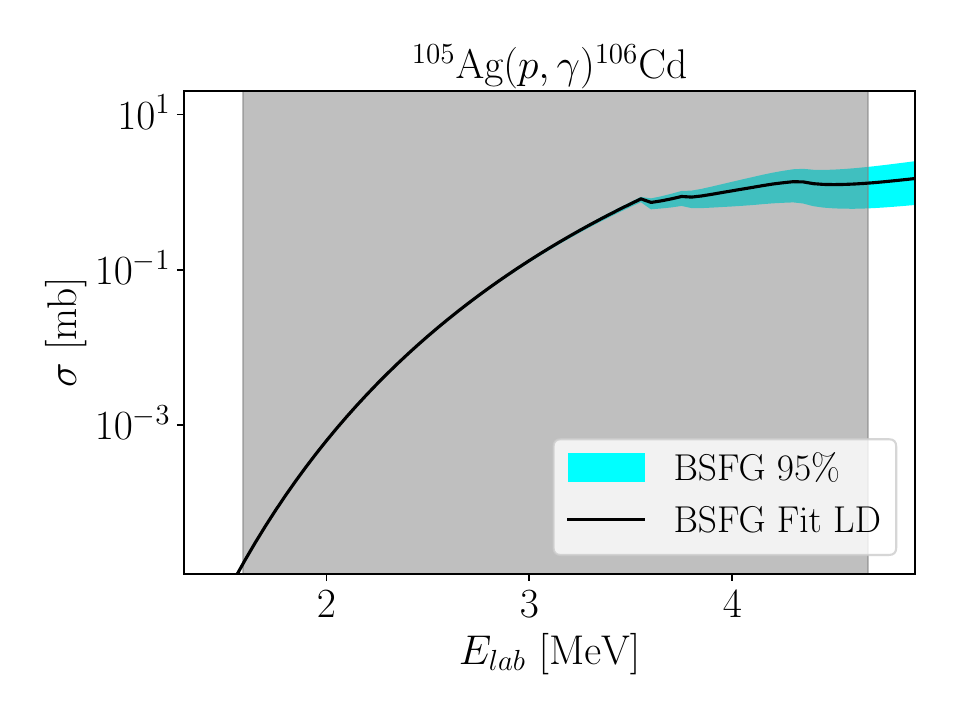}
    \caption{}
    \label{subfig: pg bsfg 106Cd}
    \end{subfigure}
     
    \caption{Calculated level densities and cross sections using the posterior distributions of the parameters $a,\delta^{BSFG}$ and $f$ of the BSFG model for the isotopes \isotope[105, 106]{}{Cd}. The best-fit value and the 95\% high density intervals obtained after Bayesian optimisation are shown with the black curve and the cyan band, respectively. The grey-shaded area represents the Gamow window for the reactions \isotope[104,105]{Ag}($p,\gamma$)\isotope[105,106]{Cd}. See text for details.}
    \label{fig: cd bfsg}
\end{figure*}

\begin{figure*}[t]
    \centering
    \begin{subfigure}{0.50\textwidth}
    \includegraphics[width=\textwidth]{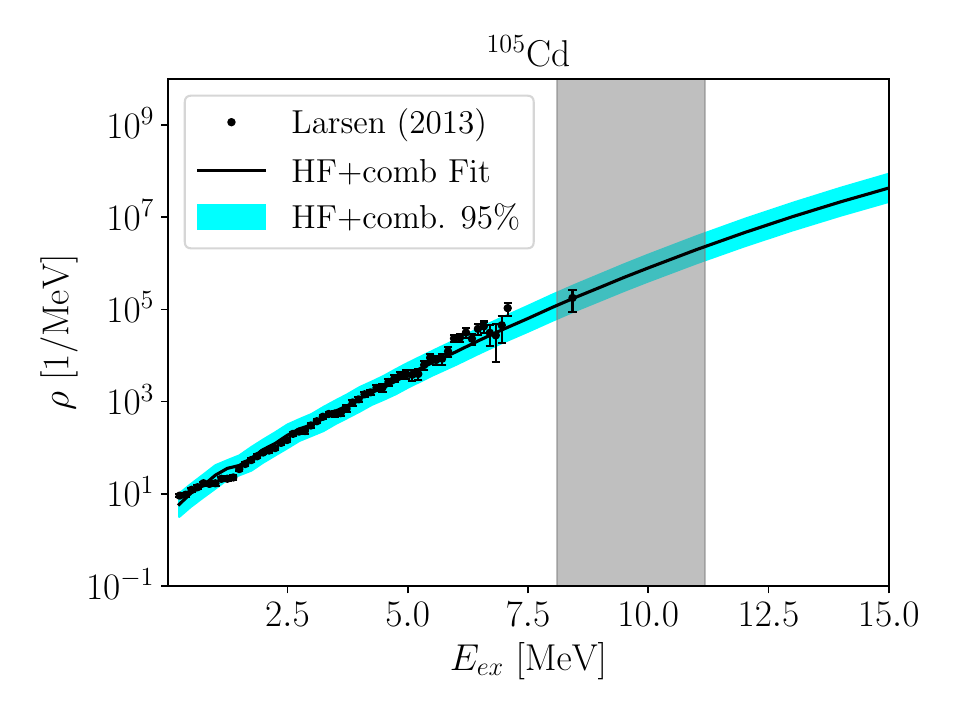}
    \caption{}
    \label{subfig: ld hfb 105Cd}
    \end{subfigure}%
      \begin{subfigure}{0.50\textwidth}
    \includegraphics[width=\textwidth]{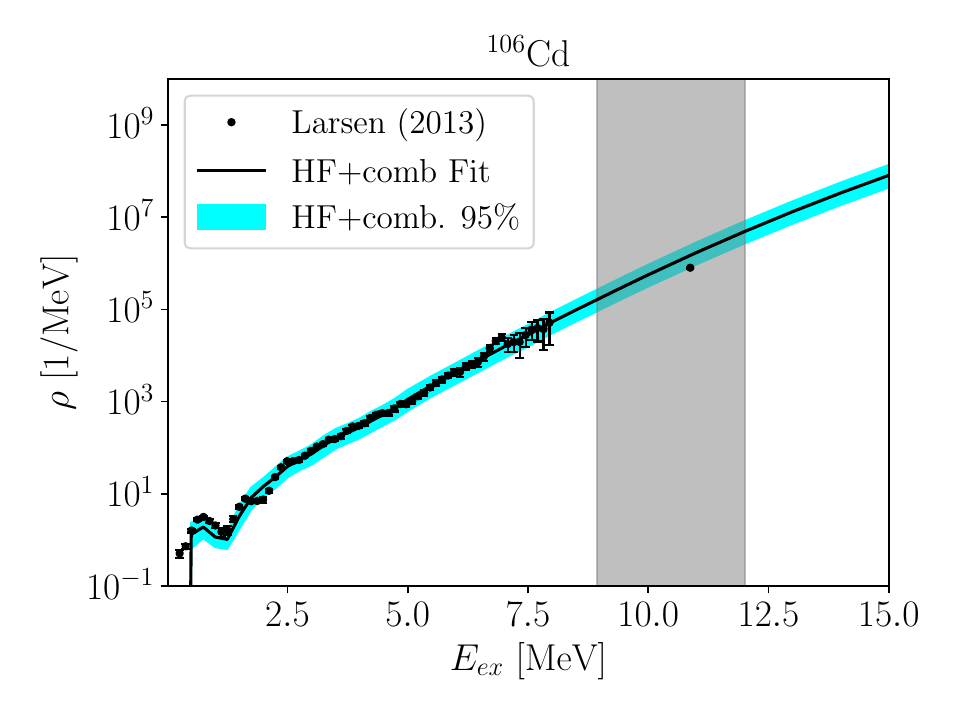}
    \caption{}
    \label{subfig: ld hfb 106Cd}
    \end{subfigure}
    \begin{subfigure}{0.50\textwidth}
    \includegraphics[width=\textwidth]{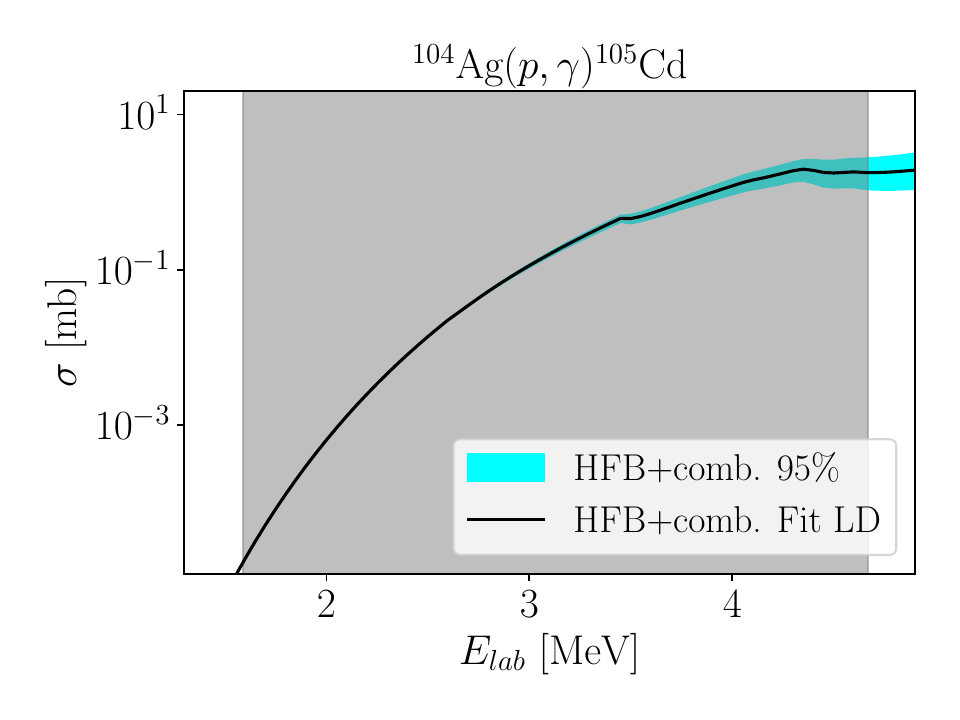}
    \caption{}
    \label{subfig: pg hfb 105Cd}
    \end{subfigure}%
    \begin{subfigure}{0.5\textwidth}
    \includegraphics[width=\textwidth]{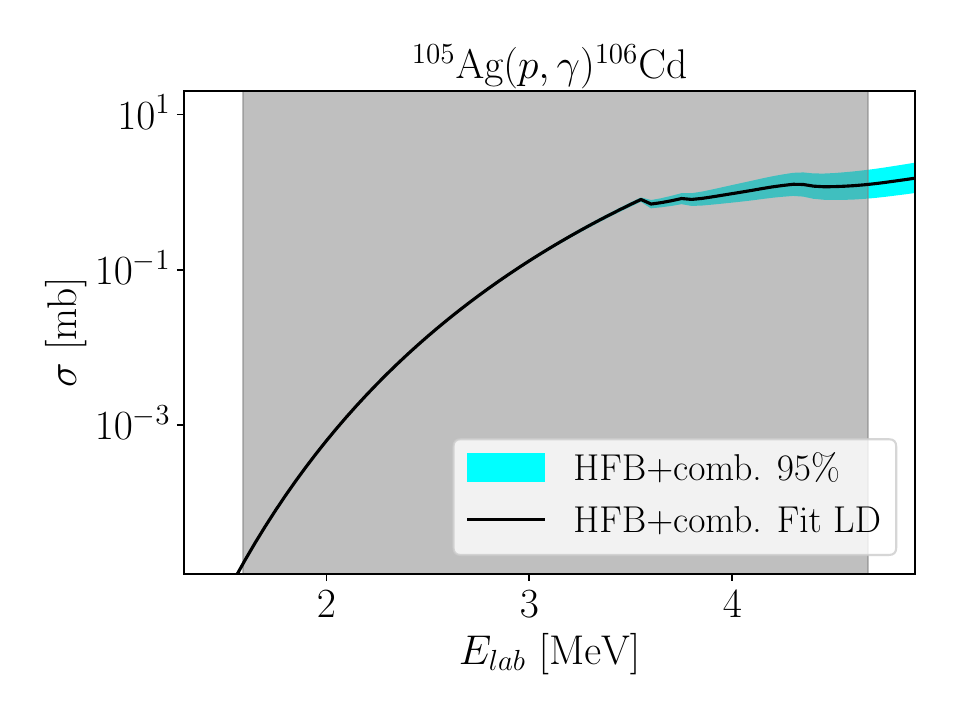}
    \caption{}
    \label{subfig: pg hfb 106Cd}
    \end{subfigure}  
    \caption{Calculated level densities and cross sections using the posterior distributions of the parameters $a,\delta^{HFB}$ and $f$ of the HFB+comb. model for the isotopes \isotope[105, 106]{}{Cd}. The best-fit value and the 95\% high density intervals obtained after Bayesian optimisation are shown with the black curve and the cyan band, respectively. The grey-shaded area represents the Gamow window for the reactions \isotope[104,105]{Ag}($p,\gamma$)\isotope[105,106]{Cd}. See text for details..} 
    \label{fig: cd hfb}
\end{figure*}

\subsection{Bayesian optimization}

Bayesian optimization is a statistical method that can optimize the free parameters of a model in order to describe a dataset. In Nuclear Reaction Theory, this approach has been successfully used for constraining OMPs for nucleon transfer reactions~\cite{Lovell_2017_PhysRevC.95.024611, Lovell_2018_PhysRevC.97.064612}. It has also been argued that the uncertainties obtained by the Bayesian approach represent more accurately the real uncertainties compared to the uncertainties obtained by frequentist approaches~\cite{King_2019_PhysRevLett.122.232502}. 

Bayesian optimization is based on Bayes' theorem:
\begin{equation}
    P(H|D ; \overline{\theta})=\frac{P(H; \overline{\theta}) P(D|H; \overline{\theta})}{P(D; \overline{\theta})}.
    \label{eq: bayes theorem}
\end{equation}
Here, $P(H ; \overline{\theta})$ is a \textit{prior} distribution, which expresses our prior hypothesis $H$ of the distributions of the model parameters $\overline{\theta}$, without the information of the dataset $D$ that we want to study. Next, $P(D|H; \overline{\theta})$ is the \textit{likelihood} function, which expresses the degree of probability that our model correctly describes the dataset, while 
$P(D; \overline{\theta})$ is the \textit{model evidence}, which expresses the sum of probabilities of all possible hypotheses. This distribution in general cannot be calculated analytically, but it is possible to sample the distribution using Markov Chain Monte Carlo (MCMC) methods. The above three distributions give the \textit{posterior} distribution $P(H|D ; \overline{\theta})$,
which expresses the probability distribution of the parameters $\overline{\theta}$ after the information obtained from the dataset.

As previously mentioned, due to the difficulty in calculating the evidence analytically, the probability distributions for the model parameters, the posterior distributions in Eq.~\ref{eq: bayes theorem} have to be sampled. For this work we have used the Goodman and Weare ensemble sampler~\cite{Goodman_2010} with the relevant software package EMCEE~\cite{Foreman_Mackey_2013}. The sampler was coupled with the TALYS code for the calculation of the theoretical level density values.

For the BSFG model, we have used as prior distributions wide gaussians of mean equal to the values derived from systematics~\cite{2007Talys}. The standard deviation of these gaussian is 100 times their mean values, in order to fully explore the parameter space (see Section~\ref{subsection: prior sens}). For the HFB+comb. model we have used priors centered at 0 and a standard deviation equal to 10, which is the limit that TALYS imposes. Table~\ref{tab: priors} summarizes all the gaussian priors and their standard deviations used for each isotope in the present work.
The likelihood is calculated from the relation:
\begin{equation}\label{eq: likelihood}
P(D|H; \overline{\theta})=  \prod_{i=1}^N \frac{1}{ \sigma_i \sqrt{2 \pi } }  \exp{ \left( -\frac{1}{2} \frac{ ( \rho^{th}_i - \rho^{exp}_i)^2 }{  \sigma_i^2 } \right) }
\end{equation}
where $\rho^{th}_i$ is the level density given by the model, $\rho^{exp}_i$ is the experimental value and $\sigma_i^2$ is calculated from~\cite{hogg2010data,emcee_doc}: 
\begin{equation}\label{eq: standard deviation}
    \sigma^2_i = \sigma_{\rho,i}^2 + (\rho_i^{th})^2 \times \exp(\ln f^2)
\end{equation}
where $\sigma_{\rho,i}$ are the measurement errors of the experimental data and the positive parameter $f$ expresses any unknown inaccuracies/noise between the models and the data. Since these inaccuracies are not known, this parameter will be optimized at the same time with the other model parameters to obtain a posterior distribution. For the parameter $f$, always a flat prior will be used between $\ln f=-10$ and $\ln f = 1$.

\section{Results and Discussion}
\label{results}

\subsection{The reactions \isotope[104,105]{Rh}($p,\gamma$)\isotope[105,106]{Pd}}

The cross sections and the relevant reaction rates of the reactions \isotope[104,105]{Rh}($p,\gamma$)\isotope[105,106]{Pd} can have an influence on the synthesis of the \textit{p} nucleus \isotope[102]{Pd}. As shown in~\cite{2006Rapp}, this nucleus is underproduced in model calculations. The aforementioned reactions are happening on the vicinity of this nucleus and their cross sections are still not measured. This is due to the difficulty in using stable beams on the unstable isotopes \isotope[104,105]{Rh}, as well as the challenge on producing the corresponding radioactive beams for inverse-kinematics measurements. Until today, the only measurements on radiative-capture reactions involving Pd isotopes have been performed in~\cite{2008spyrou,Harissopulos_2016_PhysRevC.93.025804}.  It is thus important to take advantage of the available level density data and attempt to constrain cross sections of neighboring reactions. However, the existing level density data in~\cite{Eriksen_2014_PhysRevC.90.044311} can be used to provide constraints.

The posterior distributions optimized for the level density parameters $a,\delta^{BSFG}$ of the BSFG model, as well as for the natural logarithm of the parameter $\ln f$, are shown in Fig.~\ref{fig: posteriors bsfg} and their numerical values are tabulated in Table~\ref{tab: posteriors BSFG}. As shown, the posteriors are gaussian distributions, which is expected as both the prior and the likelihood distributions are also gaussian. The 95\% high density intervals can be calculated directly from the posterior distributions and they can then be used to compare them with the level density data. The comparisons are shown in Fig.~\ref{fig: pd bsfg}a-b. together with the calculated cyan band corresponding at the 95\% of the high density intervals. A level density calculation for each isotope was also performed using the most probable value for each parameter, as given by the posterior distributions.

Using the 95\% high density intervals, the corresponding range of radiative proton-capture cross sections of the reactions \isotope[104,105]{Rh}($p,\gamma$)\isotope[105,106]{Pd} can be calculated using TALYS. The results for the BSFG model are shown in Fig.~\ref{fig: pd bsfg}c-d. 

The same calculations have been performed using the posterior distributions obtained for the parameters $c$ and $\delta^{HFB}$ of the scaling function of Eq.~\ref{eq: scaling function}, used to adjust the microscopic calculations of the HFB+comb. model to experimental data. The posterior distributions for the parameters $c$ and $\delta^{HFB}$ and $\ln f$ are shown in Fig.~\ref{fig: posteriors hfb} and in Table~\ref{tab: posteriors hfb}. The high density intervals are plotted along with the level density data for the isotopes in Fig.~\ref{fig: pd hfb}a-b and the corresponding cross sections for the reactions \isotope[104,105]{Rh}($p,\gamma$)\isotope[105,106]{Pd} in Fig.~\ref{fig: pd hfb}c-d. 

For the case of \isotope[105]{Pd}, the model uncertainties are smaller for the BSFG model by approximately factor of 2, compared to the scaled HFB+comb. model. This is not the case for the isotope \isotope[106]{Pd}, where not only the uncertainty intervals are smaller but there is a noticeably very good description of the HFB.+comb. on the lower energy parts below 3 MeV. The "oscillating" behavior in these energy range is nicely reproduced and the calculated uncertainties contain well the datasets.

When attempting to extrapolate to the Gamow Window (gray shaded area in the figures), the two models and their uncertainties can differ by approximately a factor of 4.5. In particular for the case of \isotope[105]{}{Pd}, the difference can reach even almost half an order of magnitude. The reason for this is could be traced on the different behavior of the models in higher energies. The lack of experimental level density data at energies inside the Gamow Window is also a factor, highlighting a necessity for measurements at higher excitation energies. These uncertainties are propagated to the cross sections, which are relatively smaller in low energies. This is expected, as in these energies, the cross sections depends mainly on the OMPs and not on the level density.

\begin{table}[]
    \centering
    \caption{Posterior distributions for the parameters $a, \delta^{BSFG}$ and $\ln f$ of the BSFG model. The most probable values $a_{MP}, \delta_{MP}^{BSFG}$ and $(\ln f)_{MP}$ and the 95\% high density intevals $(a_1,a_2), (\delta_1^{BSFG},\delta_2^{BSFG})$ and $(\ln f_1,\ln f_2)$ are given.}
    \begin{tabular}{c|c|c|c|c}
    Isotope & \isotope[105]{Pd}  & \isotope[106]{Pd} & \isotope[105]{Cd} & \isotope[106]{Cd} \\ \hline  
    $a_{MP}$ & 12.32 & 11.83 & 10.98 & 11.19   \\ 
    $a_1$ & 12.03 & 11.54 & 10.67 & 10.9    \\ 
    $a_2$ & 12.61 & 12.12 & 11.28 & 11.38    \\ 
    $\delta_{MP}^{BSFG} $ & -0.97 & -1.00 & -0.91 & -0.95 \\ 
    $\delta_1^{BSFG}$ & -1.07 & -1.08 & -1.02 & -1.03\\
    $\delta_2^{BSFG}$ & -0.90 & -0.88 & -0.78 & -0.85 \\
    $(\ln f)_{MP}$ & -1.93 & -0.99 & -1.51 & -1.00 \\
    $\ln f_1$ & -2.23 & -1.18 & -1.77 & -1.20 \\
    $\ln f_2$ & -1.58 & -0.74 & -1.13 & -0.75
    \end{tabular}
   
    \label{tab: posteriors BSFG}
\end{table}

\begin{table}[]
    \centering
    \caption{Posterior distributions for the parameters $c, \delta^{HFB}$ and $\ln f$ of the HFB+comb. model. The most probable values $c_{MP}, \delta_{MP}^{HFB}$ and $(\ln f)_{MP}$ and the 95\% high density intevals $(c_1,c_2), (\delta_1^{HFB},\delta_2^{HFB})$ and $(\ln f_1,\ln f_2)$ are given.}
    \begin{tabular}{c|c|c|c|c}
    Isotope & \isotope[105]{Pd}  & \isotope[106]{Pd} & \isotope[105]{Cd} & \isotope[106]{Cd} \\ \hline  
    $c_{MP}$ & -0.31 & -0.31 & -0.02 & 0.51   \\ 
    $c_1$ & -0.42 & -0.39 & -0.1 & 0.47    \\ 
    $c_2$ & -0.13 & -0.23 & 0.09 & 0.56    \\ 
    $\delta_{MP}^{HFB}$ & -1.13 & 0.14 & -0.54 & 0.36 \\ 
    $\delta_1^{HFB}$ & -1.23 & 0.05 & -0.63 & 0.32 \\
    $\delta_2^{HFB}$ & -0.85 & 0.22 & -0.39 &  0.39 \\
    $(\ln f)_{MP}$ & -1.76 & -1.38 & -1.74 & -1.35 \\
    $\ln f_1$ & -2.08 & -1.59 & -2.05 & -1.58\\
    $\ln f_2$ & -1.49 & -1.09 & -1.41 & -1.06
    \end{tabular}
   
    \label{tab: posteriors hfb}
\end{table}

\subsection{The reactions \isotope[104]{Ag}($p,\gamma$)\isotope[105]{Cd} and \isotope[105]{Ag}($p,\gamma$)\isotope[106]{Cd}}

The Cd isotopic chain hosts a \textit{p} nucleus (\isotope[106]{Cd}) whose abundance is significantly underproduced by astrophysical models~\cite{2006Rapp}, as in the case of \isotope[102]{}{Pd}. Proton-capture cross sections studies reaching \isotope[108]{}{Cd} are more accessible due to the existence of the stable target \isotope[107]{}{Ag}. These cross sections have been studied extensively in~\cite{Khaliel_2017_PhysRevC.96.035806, Heim_2020_PhysRevC.101.035805} and have sufficiently constrained the cross section ranges inside the Gamow window.

On the other hand, for the case of \isotope[105,106]{}{Cd}, similar data are non-existent because of the lack of stable targets. In~\cite{Rauscher2013}, theoretical calculations predict high degree of competition between the reaction \isotope[104]{Ag}($p,\gamma$)\isotope[105]{Cd} and \isotope[105]{Ag}($n,\gamma$)\isotope[106]{Cd}. It is thus necessary to take advantage of the level density data for \isotope[105,106]{}{Cd} and attempt to constrain the range of cross sections of the proton capture reactions \isotope[104]{Ag}($p,\gamma$)\isotope[105]{Cd} and \isotope[105]{Ag}($p,\gamma$)\isotope[106]{Cd}, based on the posterior distributions sampled by optimizing the BSFG and HFB+comb. models on the OSLO data for these isotopes~\cite{Larsen_2013_PhysRevC.87.014319}. 

The same procedure is followed as for the previously described Pd isotopes. The posterior distributions of the BSFG model parameters $a, \delta^{BSFG}$ and $\ln f$ are shown in Fig.~\ref{fig: posteriors bsfg} for the isotopes \isotope[105,106]{Cd}. Furthermore, the posteriors of the scaled HFB+comb. model are also shown in Fig.~\ref{fig: posteriors hfb} for the parameters $c, \delta^{HFB}$ and $\ln f$. 

The calculated level densities and the cross sections of the reactions \isotope[104,105]{Ag}($p,\gamma$)\isotope[105,106]{Cd} are compared with the data in Fig.~\ref{fig: cd bfsg} for the BSFG model 
and in Fig.~\ref{fig: cd hfb} for the HFB+comb. model. The range of uncertainties is significantly smaller for the HFB+comb. model in both cases, as well as the description of the low-energy data. These results point out that uncertainties can significantly depend on the choice of the model.

\subsection{Prior Sensitivity and autocorrelation}
\label{subsection: prior sens}
It is useful for every study using Bayesian inference to include a sensitivity study of the posterior distribution due to the prior distribution. An efficient way to check this sensitivity is described in~\cite{Lovell_2017_PhysRevC.95.024611}, and consists of checking the posterior distribution by varying the standard deviation of the Gaussian priors. 

\begin{figure}[t]
    \centering
    \includegraphics[width=\columnwidth]{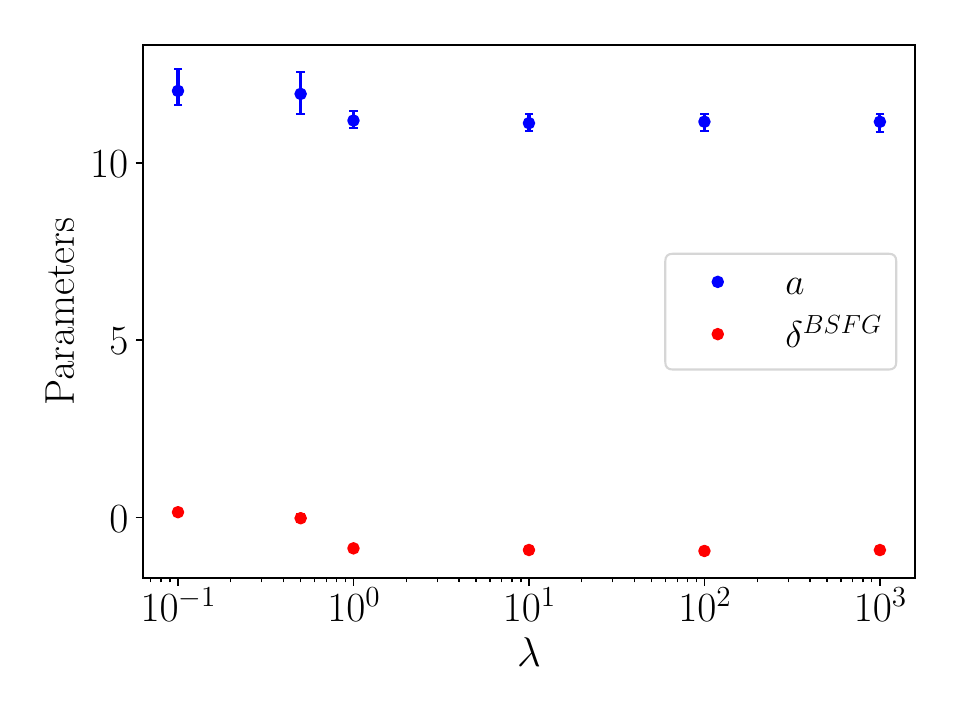}
    \caption{Prior sensitivity of posterior distributions the parameters $a$ and $\delta^{BSFG}$ of the BSFG for the isotope \isotope[106]{Cd}. The standard deviation of the gaussian prior is given by the product of the factor $\lambda$ and the mean of the gaussian prior. For large values of $\lambda$, the prior distributions becomes more wide (uninformative prior), stabilizing the posterior to specific values. Error bars represent the 95\% high density intervals of the posteriors. See text for details.}
    \label{fig: prior sensitivity}
\end{figure}

\begin{figure}[t]
    \centering
    \includegraphics[width=\columnwidth]{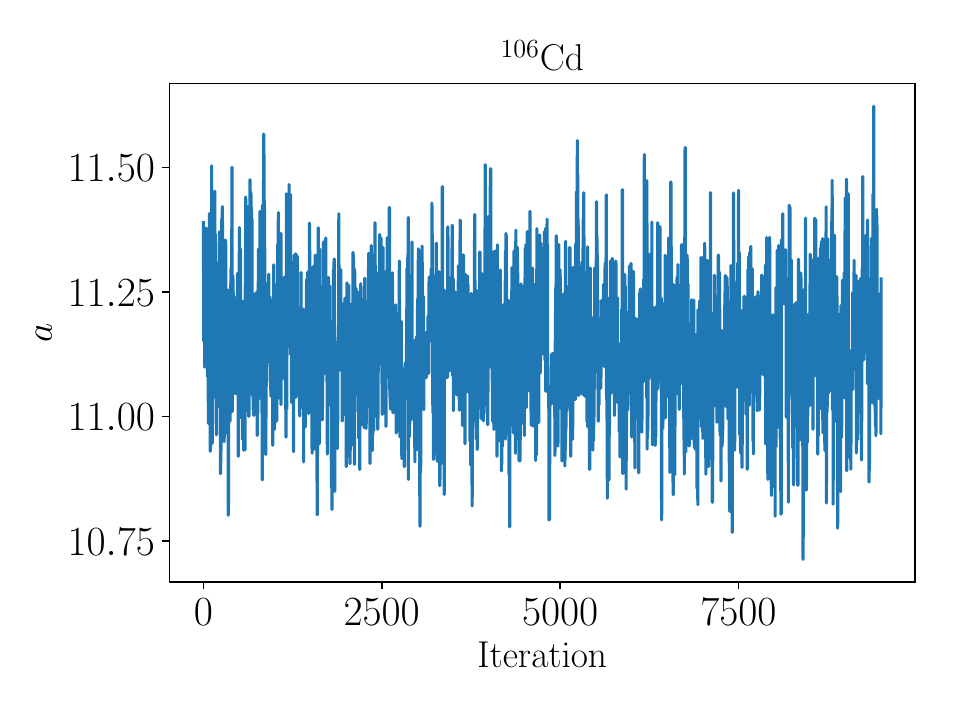}
    \caption{Trace of the parameter $a$ of the BSFG for the nucleus \isotope[106]{Cd}.}
    \label{fig: trace}
\end{figure}

In Fig.~\ref{fig: prior sensitivity}, the posterior 95\% high density intervals for the parameters $a$ and $\delta^{BSFG}$ of the BSFG model are shown as a function of the standard deviation of their gaussian priors. 
The sensitivity check starts from a gaussian distribution with means, $\mu$, equal to the systematic values of level densities and a standard deviation is equal to $\lambda \times a_M, \delta_M^{BSFG}$ for $a$ and $\delta^{BSFG}$, respectively. The standard deviation of the prior can then be varied by increasing the factor $\lambda$, that is by increasing the width of the gaussians. As shown in Fig.~\ref{fig: prior sensitivity}, for $\lambda<1.$, there seems to be a strong variation of the posterior as a function of the standard deviation of the prior. After this threshold, the posteriors seem to be the same, something that is expected as the prior becomes too uninformative. In all calculations presented in this work, the factor $\lambda$ is chosen large enough $(\lambda = 100)$ in order to minimize prior influence and fully explore the parameter space, as argued in~\cite{King_2019_PhysRevLett.122.232502}.

The memory effect (autocorrelation) of the Markov Chain can be solved by running a large chain or thinning the chain by a step. In Fig.~\ref{fig: trace}, the trace of the parameter $a$ of the BSFG is shown for the isotope \isotope[106]{Cd}. A chain of 10,000 events was run using six walkers. As shown, no correlating structure is viewed in the trace. 

\begin{figure}[t]
    \centering
    \includegraphics[width=\columnwidth]{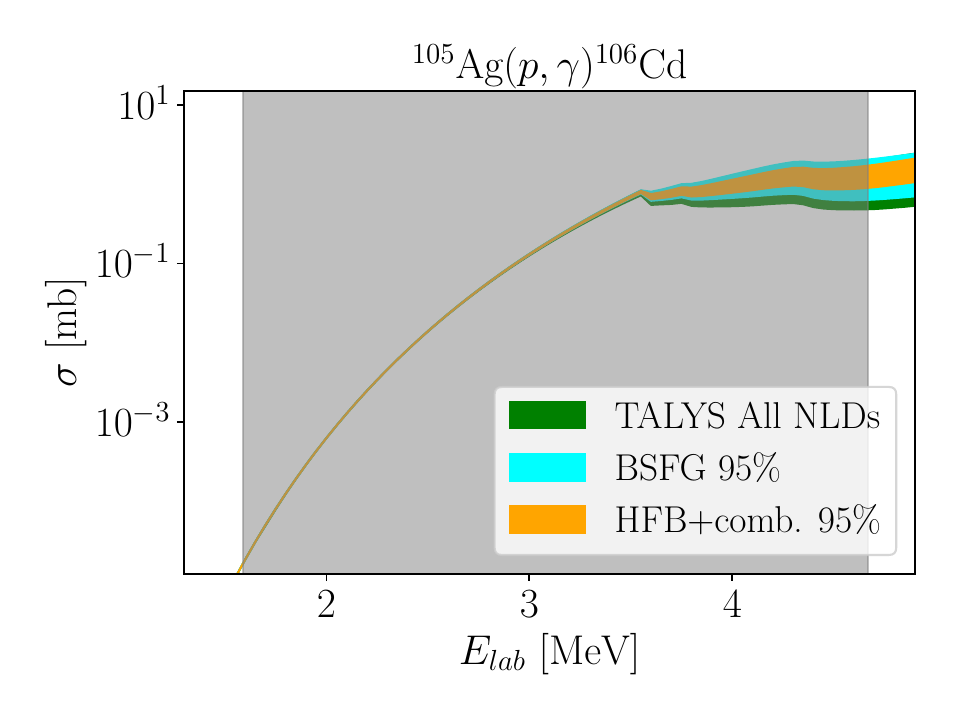}
    \caption{Cross sections calculations comparing the range of all six default TALYS level density models with the Bayesian 95\% high density intervals of the BSFG and HFB+comb. models for \isotope[106]{Cd}. See text for details.}
    \label{fig: comp all}
\end{figure}

\subsection{Comparison with all TALYS models}

A comparison with the range of all available default TALYS models has also been performed, in order to compare the ranges of cross sections for the case of reaction \isotope[105]{Ag}($p,\gamma$)\isotope[106]{Cd}. The results are compared with the 95\% confidence levels that were calculated using the level density data and the BSFG and HFB+comb. models. The comparison is shown in Fig.~\ref{fig: comp all}.

The range between the maximum and minimum of all 6 default TALYS models can cover a relatively large range which can reach a difference of 1.5 mb for the higher energies in the Gamow Window. With the use of level density data, this uncertainty interval is significantly reduced to approximately 0.6 mb for the HFB+comb. model. This highlights the importance of the level density data in constraining important cross sections for the \textit{p} process. It has to be noted however that extrapolation to the Gamow window is necessary, as there are usually there are no level density data at such excitation energies. This can pose significant challenges in constraining the cross sections and highlights the need for both level density and cross section data inside the Gamow window.

\section{Conclusions and Future directions}
\label{conclusions}

Four important radiative proton-capture reactions within the \textit{p} process reaction have been studied by using Bayesian inference in order to quantify the uncertainties that arise from measured level density data. The 95\% high density intervals using the normalized level density data from the OSLO database \cite{OSLO_data} have been calculated and propagated to the $(p,\gamma)$ cross sections. While the present study is oriented on the cross sections of proton-capture reaction of astrophysical interests, uncertainties in level densities could also constrain other observables in different domains such as in fission~\cite{PIAU2023137648}.

In three out four cases studied in the present work, the HFB+comb. model seems to provide smaller high density intervals. An exception seems to be the case of \isotope[105]{}{Cd}, where uncertainties where found smaller by using the BSFG model. This indicates that different models can give different results when optimized on the same dataset, making the choice of the model important. In all cases though, the shape of the high density intervals of the HFB+comb. model seem to describe very well the low-energy level densities.

The level densities are not the only factor that affects the final proton-capture cross section. The optical model potential and the $\gamma$-strength function have also a significant impact. While the proton-nucleus optical models are well constrained~\cite{Koning2003}, there are still uncertainties arising from the different $\gamma$SF models. A Bayesian approach has already been used for the $\gamma$SF for the reaction \isotope[93]{}{Nb}$(p,\gamma)$\isotope[94]{Mo}, by using the OSLO data. In combination with the present work on level densities, uncertainty quantification on $\gamma$SF models could give as a more complete picture on the cross sections uncertainties.

While this work is focused on proton-capture reactions, the impact of $\alpha$-captures is also significant for the \textit{p} process network. In this case there exists also $\alpha$-nucleus potential problem, which is not as well constrained as the nucleon-nucleus potentials. A Bayesian study on existing data could give insight on the uncertainties related to the $\alpha$ optical potential, which is important in particular for the lowest part of the Gamow Window.

\section*{Acknowledgements}
The authors are grateful to the LABEX Lyon Institute of Origins (ANR-10-LABX-0066) Lyon for its financial support within the Plan France 2030 of the French government operated by the National Research Agency (ANR). A. Chalil is grateful to Dr. Valerie Lapoux, for the useful discussions over the present work.

\bibliography{ld_uncertainties.bib}
\bibliographystyle{apsrev}


\end{document}